# Cognitive Aging as Interplay between Hebbian Learning and Criticality

Sakyasingha Dasgupta

*Bernstein Center for Computational Neuroscience, III Institute of Physics,*
*University of Göttingen, Göttingen, 37077, Germany*
dasgupta@physik3.gwdg.de

## ABSTRACT

Cognitive ageing seems to be a story of global degradation. As one ages there are a number of physical, chemical and biological changes that take place. Therefore it is logical to assume that the brain is no exception to this phenomenon. The principle purpose of this project is to use models of neural dynamics and learning based on the underlying principle of self-organised criticality, to account for the age related cognitive effects. In this regard learning in neural networks can serve as a model for the acquisition of skills and knowledge in early development stages i.e. the ageing process and criticality in the network serves as the optimum state of cognitive abilities. Possible candidate mechanisms for ageing in a neural network are loss of connectivity and neurons, increase in the level of noise, reduction in white matter or more interestingly longer learning history and the competition among several optimization objectives. In this paper we are primarily interested in the affect of the longer learning history on memory and thus the optimality in the brain. Hence it is hypothesized that prolonged learning in the form of associative memory patterns can destroy the state of criticality in the network. We base our model on Tsodyks and Markrams [49] model of dynamic synapses, in the process to explore the effect of combining standard Hebbian learning with the phenomenon of Self-organised criticality. The project mainly consists of evaluations and simulations of networks of integrate and fire-neurons that have been subjected to various combinations of neural-level ageing effects, with the aim of establishing the primary hypothesis and understanding the decline of cognitive abilities due to ageing, using one of its important characteristics, a longer learning history.

*"It may well be there is something else going on in the brain that we don't have an inkling of at the moment."* - Physicist Roger Penrose

# Chapter 1

# Introduction

As indexed by performance in various memory tasks and extensive studies carried out over the last two decades, it is seen that cognitive functionalities show a marked decline as a function of the advancing age. Some of the cognitive effects typical in ageing are the slow-down of processing speed, decline in the ability to encode new sequences of memory and facts and the degradation of selective attention or the decline of various types of memory. The question of how best to account for this decline is a long standing one. Furthermore, researchers from cognitive neuroscience, cognitive psychology as well as computational neuroscience have extensively debated whether the decline is attributable predominantly to a deficit in the initial encoding of information into memory or in its subsequent retrieval.

## 1.1 Motivation

Interestingly there have been numerous neuronal models that have been introduced in order to account for ageing in correspondence with known physiological changes in the brain. However it is a known fact that ageing of the brain goes hand in hand with a continuous learning process. There is constant encoding and decoding of stored or new memories and associations of stored or retrieved patterns, hence although increase or decrease in neurons [8], changes in the brain white matter or increase in noise in the brain [12], emerge as suitable candidates to explain the cognitive decline, *a longer learning* *history* [13] appeals to me as one of the most ideal candidate to lay the basis for a neuronal model to explain the ageing brain.

Although Babcock (2002) showed that the performance in psychological tests tend to decrease with age, Salthouse (1999) [7] had achieved strongly contrasting results, by showing performance improvements in everyday situations. Laying my basis on these observations I believe that learning in neural networks can serve as a suitable model for the acquisition of skills and knowledge in early development stages of the brain. Learning in this sense can be seen to follow an optimization paradigm to maintain stability or a state of *criticality* in the brain that serves as the optimal state for speed and comprehensive trade off between a large set of combinations of features. Furthermore the sensitivity and dynamic range of a network made of neuron-like elements is now shown to be maximized at the critical point of a phase transition [15]. Hence based on the assumption that this optimization paradigm remains valid for later stages i.e. the ageing process, we explore neural models of cognitive ageing keeping a longer learning history as our primary basis.

More than often the brain is depicted as a compartmentalised box, with different sections of the brain aiding in varying or overlapping functionalities. Contrary to this view I tend to think of the brain as a natural complex dynamic system with an inherent mechanism favouring self-organization. In this regard Self-organized criticality [1] is essentially one of the key concepts to describe the emergence of complexity in natural systems and hence forms and integral part of this project. Furthermore the presence of critical behaviour in neuronal systems have been proven both analytically [4, 5] as well as using experimental procedures by Beggs and Plenz [2, 22]. Although initial research required the parameters of the network to be fine-tuned externally to a critical state, by assuming dynamical synapses in a spiking neural network, the neuronal avalanches turn from an exceptional phenomenon into a typical and robust self-organized critical behaviour. Basically multiple dynamical regimes can be reconciled with parameter-independent criticality. This along with the concept of associative recall from the standard Hopfield model [14] forms the prime motivation of the project.

## 1.2 Aim and important question

The aim of the project is to provide a simulated evaluation of neuronal models of dynamic synapse, to explore the degradation of the cognitive functions with age (i.e. in this context, of learning new patterns of memory) using an underlying paradigm of associative memory with dynamical synapses, combining Hebbian learning and the integrate fire neuron variant of the Hopfield model in the context of the phenomenon of Self-organized criticality. I would like to emphasize here that although I am considering a highly simplified model, I believe "***universality***" and "***criticality***" might be features that still permit a generalization to the real brain (in this case concerning the cognitive effects of ageing). In part this thesis tests exactly this claim making a small but important contribution towards a understanding of the possible organisation of memory. As a final goal the project not only is another essential contribution towards the understanding of criticality in the brain and its implications on the ageing process, but in general help in using neural network models to understand the intrinsic chaotic behaviours of the brain during epileptic seizures (i.e. equated with excessive neural activity caused in the brain due to neuronal avalanche along with the loss of criticality due to prolonged learning).

Essentially the following questions or ideas are explored:

- What are the possible outcomes of a combination of the standard Hebbian learning rule and the concept of Self-organized criticality?
- Establish the original hypothesis (as mentioned above) using simulations of network of integrate fire neurons and depict the destructive interference of Hebbian learning and Self-organized criticality.
- Explore the concept of dynamic synapses and associative memory.
- Whether learning influences parameters like the overlap of memory states (*m*) and the mean squared deviation from power-laws ($\Delta \gamma$) causing dilution of the network?

## 1.3 Implementation

Having understood the various cognitive effects of ageing specifically the implications on memory and general intelligence (g), models of neural dynamics and learning are used to account for these age related cognitive effects. Based on the original hypothesis, a neural model that allows associative recall is considered, drawing inspiration from the stochastic Hopfield model. A simulation of learning in neural networks with dynamical synapses, which are found to robustly exhibit critical behaviour, is implemented. However it is important to note that though for the purpose of understanding the ageing process better, extensive research was carried out on the effect of ageing on general intelligence, the neural models thus devised don't make any claims to model intelligence or form the basis to calculate *g*.

As a starting point with slight modifications to the model of Levina et. al [25] using dynamical synapses in neural networks , self-organized criticality is observed and the necessary observables are noted. Having achieved this Hebbian learning is combined with the previous model and dynamic synapses with associative memory is explored. Furthermore synaptic adaptation beyond Hebbian dynamics are also explored, in order to account for realistic regulations occurring in the brain, that help in maintain the state of criticality.

It is believed that the critical state makes provision for optimum speed and comprehensive trade-off in the exploration of a large set of combinations of features. An experimental prediction of the effect of long term learning using a model of Hebbian learning of network connectivity and exposure to various input patterns is carried out. Both random as well as controlled orthogonal patterns are tested and their effect on criticality is observed. Due to the space time complexity of the project a moderate sized network consisting of 200 to 500 neurons are used. As mentioned previously although I consider simplified network dimensions, the obtained results can well be generalized to biological realms.  The later chapters will discuss the essential models and results obtained in greater depth.

# Chapter 2

# Theoretical Background

Age and change seemingly go hand in hand. As one ages there are a number of physical, chemical or biological changes that take place. Therefore it is logical to assume that the brain is no exception to this phenomenon. Although no two individuals are the same, general trends of age related degradation of cognitive functions exist. There are various structural, chemical and genetic changes that occur in old age. Some changes in the ability to think are considered a normal part of the ageing process. Abilities such as encoding new memories of episodes and facts, working memory [6], and processing speed are relatively impaired and tend to decline in old age [7].

Interestingly there can be a change in **General Intelligence** with age, though the rate may vary depending on the type namely **Fluid Intelligence** (mechanics) or **Crystalline Intelligence** (pragmatics). In generic terms Fluid intelligence is the ability to make sense out of confusion and solve new problems. It is the ability to draw inferences and understand the relationships of various concepts, independent of acquired knowledge. It is integral to all logical problem solving, especially scientific, mathematical and technical problem solving. Crystalline Intelligence is the ability to use skills, knowledge and experience. Though it cannot be directly equated with memory, it does rely on accessing information from long-term memory. Furthermore working memory [45] capacity is believed to be closely related with fluid intelligence and may potentially be the reason for individual differences in general intelligence.

Psychologists have observed that there is a global increase in performance during youth, whereas during ageing, tasks that can be discriminated by demanding skills or **mechanics** (Figure 1.1 (a)) decrease with age. However those tasks requiring **pragmatics** (Figure. 1.1 (b)) don't decrease but gradually saturate with age. This leads to an interesting observation that young adults often demonstrate high levels of flexibility, where as the aged tend to become wiser i.e. attain expert knowledge.

[8]

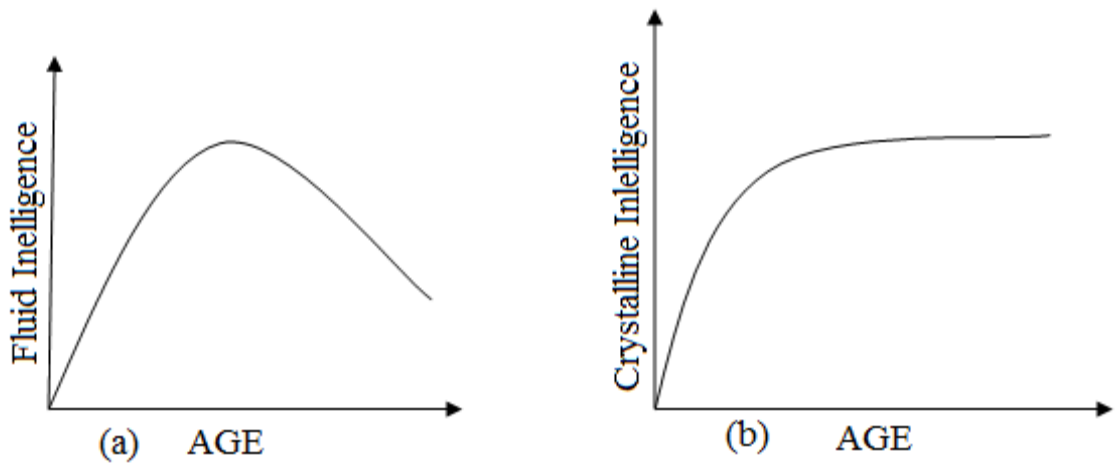

*Figure 1.1*: *(a) The change in fluid intelligence with the ageing process is depicted. It peaks in adulthood, and then gradually decreases with growing age. (b) The change in crystalline intelligence with the ageing process is depicted. As on grows there is a life-long improvement in crystalline intelligence and it gradually saturates.*

Over the years several parameters of neural models have been introduced in order to account for ageing in correspondence to known physiological changes in the brain. Shefer(1973) [8] built a model of aging using the reduction of the number of neurons as the degradation parameter. Horn and Cattel [9] accounted for the cognitive effect of aging, modelling the reduction of brain white matter by tweaking the number of neural synapses. Furthermore Brown et al. [11] and Braver et al. [10] used neural models to depict the reduction of cognitive control and suboptimal neuro-modulation respectively. More recently Li et al. [12] presented a neuro-computational model of the increase in noise (Stochastic Resonance) in the brain with age. Nadel and Moscovitch [13] talked about longer learning history in their 1997 paper on memory consolidation.

Now having provided an insight into the concept of ageing, memory and the learning process, in the following sections I briefly introduce and review few of the concepts and models most crucial to his project. The primary purpose is to introduce the models and broadly describe the areas on which work has already been done to considerable depth. Detailed descriptions of models that support the present work are provided in Section 2.1.

## 2.1 Hebbian Learning and the Hopfield Model

In his 1949 book *The Organization of Behaviour*, Donald Hebb predicted a form of synaptic plasticity with the following property:

*When an axon of cell A is near enough to excite a cell B and repeatedly or persistently takes part in firing it, some growth process or metabolic change takes place in one or both cells such that A's efficiency, as one of the cells firing B, is increased.*

This is more commonly known as the standard Hebb's Rule or Hebb's postulate. In general **Hebbian theory** describes a basic mechanism for synaptic plasticity where in an increase in the synaptic efficacy arises from the presynaptic cell's *repeated* and *persistent* stimulation of the postsynaptic cell. The theory is often summarized as "*cells that fire together, wire together*". From the point of view of artificial neurons and artificial neural networks, Hebb's principle can be described as a method of determining how to alter the weights between model neurons. In General term the weight between two neurons increases if the two neurons activate simultaneously; and reduces if they activate separately.

This can be described in formula by:

$$w_{ij} = \frac{1}{N} \sum_{k=1}^{N} x_i^k x_j^k \quad , \tag{1.1}$$

where $w_{ij}$ is the weight of the connection from neuron *j* to neuron *i*, *p* is the number of training patterns, and $x_i^k$ the $k^{th}$ input for neuron *i*. This is learning by epoch (weights updated after all the training examples are presented).

A number of theorists have formulated neural network models with the goal of explaining how Hebbian synaptic plasticity could be used to store memories. For the purpose of this project and even historically the most significant of these is the 1982 paper by John J. Hopfield **"Neural Networks and Physical Systems with Emergent Collective Computational Abilities"** introduced the Hopfield network a recurrent neural network for modelling associative memory. Hopfield networks serve as content-addressable memory systems with binary threshold units. In general a associative memory model using the Hebb rule (1.1) for all possible pairs *i j,* with binary units and asynchronous updating is usually called a **Hopfield Model**.

[10]

Although there existed many ingredients of this model prior to the publication of this historical paper, Hopfield [14] brought all of them together, introduced an energy function and emphasized the idea of stored memories as ***dynamical attractors***.

## 2.1.1 The Associative memory problem

Associative memory is the "fruit fly" or the "Bohr atom" problem of this field. It illustrates in about the simplest possible manner the way that collective computation can work.

In general the associative memory emphasizes the following problem:

*Store a set of p patterns $\xi_i^\mu$ in such a way that when presented with a new pattern $\zeta_i$, the network responds by producing whichever one of the stored patterns most closely resembles $\zeta_i$.*

Suppose the Hopfield network is exposed to P binary patterns $\xi_i^1......\xi_i^P$ during a training phase. More specifically, the state $S_i$ is set to each pattern in turn, and the synaptic weights are changed by the Hebbian update. If $W_{i,j} = 0$ at the beginning of the training phase, Hebbian learning will result in:

$$W_{i,j} = \frac{1}{N} \sum_\mu \xi_i^\mu \xi_j^\mu \qquad (1.2)$$

The pre-factor 1/N corresponds to a particular choice of learning rate $\eta = \frac{1}{N}$. One of the most important contributions of the Hopfield [14], 1982 paper was the introduction of the idea of an *energy function* into neural network theory. Hence for the network considered above the energy function H is,

$$H = -\frac{1}{N}\sum_{i,j} w_{i,j} s_i s_j \tag{1.3}$$

The double sum is over all i and all j. The i = j terms are of no consequence because $S_i^2 = 1$; they just contribute a constant to H, and in any case we could choose $W_{i,j} = 0$. The energy function is a function of the configuration $\{S_i\}$ of the system, where $\{S_i\}$ means the set of all the $S_i$'s.

The most remarkable property of the energy function is that *it always decreases (or remains constant) (or has to be minimised to find a stable optimum state) as the system evolves according to its dynamical rule.* Thus one can say that the memorised patterns or attractors are at the local minima of the energy surface. Hence it can be said that there is an underlying *Lyapunov function* for the activity *dynamics*.

This gives the Hopfield model the capability to perform not only a recognition or retrieval task, but it can also be seen as solving an optimization problem i.e. the system is expected to find a configuration which minimises an energy function. In the context of ageing this forms as a basis for understanding associative memory of random uncorrelated patterns which would in turn shed light upon the loss of memory as a cognitive effect of ageing.

Having said this, it is imperative to mention that due to the underlying simplicity introduced, the standard Hopfield model does suffer from certain limitations, some of which relevant to this work have been improved upon, in order to build a more robust model.

- The standard Hopfield model restricts the neurons to either an active (+1) or a passive (-1) states. Hence for an *N*-neuron network with *P*-patterns, $\xi_i^\mu = \pm 1$ with equal probability. This implies that during the retrieval process 50% of the neurons are active, on average. This does not correlate with the neurophysiological evidence, which indicates that the mean firing rates are significantly lower than 50%. This is improved upon in this work by considering a Hopfield model using integrate and fire neurons which is biologically more relevant (Section 3.2).

- The Hopfield dynamics and energy function both have a perfect symmetry. i.e. $(S_i \rightarrow -S_i), \forall i$. Hence with every stored pattern the network also stores the reversed patterns. Furthermore using the standard Hebb's rule (1.1) the patterns are uncorrelated; namely in a large network,

$$\frac{1}{N}\sum_i \xi_i^\mu \xi_i^\nu = 0, if \mu \neq \nu .$$  (1.4)

This result in the presence of local minima, giving rise to spurious states or spurious minima [33], and therefore the memory is seen to not work perfectly. There are a number of tricks that can be considered for this, including Amit et. al [33] concept of correlated biased patterns. In this regard we make use of a modified Hebbs rule [34, 35]

$$W_{ij} = \frac{1}{Nf(1-f)} \sum_{\mu=1}^{P} (\xi_i^\mu - f)(\xi_j^\mu - f),$$  (1.15)

This represents an optimal learning rule for associative memory networks.

- From the stand point of this thesis, as well as biologically another drawback of the standard Hopfield model is the consideration of synapses as single numbers as opposed to dynamical synapses. A considerable amount of numerical and theoretical studies have revealed that a large population of excitatory neurons with depressing synapses exhibit complex regimes of activity [22, 5, 25]. Furthermore Senn et al. [40] in 1998 showed that synaptic depression could also serve as a mechanism provoking rhythmic activity and hence be important to pattern generation. Although effects of synaptic dynamics on complex neural functions like associative memory are still not completely established, drawing inspiration from the works of Bibitchkov, Herrmann & Geisel [31] we instead of the fixed points of the Hopfield network consider associative memory with dynamic synapses. This is explained further in Section 3.1. and Section 3.2.

## 2.2 Self-Organized Criticality in Neural networks

In physics, a critical point is a point at which a system changes radically its behaviour or structure, for instance, from solid to liquid. In standard critical phenomena, there is a control parameter which an experimenter can vary to obtain this radical change in behaviour. In the case of melting, the control parameter is temperature. Self-organized critical phenomena, by contrast, is exhibited by driven systems which reach a critical state by their intrinsic dynamics, independently of the value of any control parameter. According to the concept a system is believed to *self-organize* into a critical state where the system observables are distributed according to a power-law (**Appendix 1**). The combination of dynamical minimal stability and spatial scaling leads to a power law for temporal fluctuations. The noise propagates through the scaling clusters by means of a "*domino*" effect upsetting the minimally stable states.

In general the term "Self-organized criticality" emphasizes two aspects of the system behaviour. *Self-organization* is used to describe the ability of certain non-equilibrium systems in the absence of control or manipulation by an external agent to develop specific structures and patterns. The word *criticality* as mentioned above is used to emphasize the similarity with phase transitions - where in a system stays at the borders of stability and chaos such that the response of the system to an external perturbation is non-linear and even small disturbances can lead to avalanches of all sizes. The concept was originally coined and introduced by Bak et al. [1] who proposed the very first SOC model in order to explain the widespread occurrences of temporal and spatial power-law scaling in nature. Moreover it was observed that many non-equilibrium systems, composed of a large number of interacting components and driven by an external force, evolve towards critical state through interaction of the constituents.

## 2.2.1 Intuitive Understanding: The Sand-pile model

A good metaphoric picture is that of a pile of sand onto which sand is poured at a very slow rate. Initially, the pile is flat and stable. However, with the continuity of the process, the pile becomes steeper and small avalanches will occur. Eventually there will be avalanches of all sizes, up to the full size of the system. This is when the self-organized critical state is believed to have been reached, with most of the surface very steep. As a result even slight alterations in the configuration of the sand-pile can cause large changes in the flow of sand.

The mathematical modelling of sand-piles bears striking resemblance to models of neuronal functions. In sand models, each site receives sand grains from neighbouring sites. Comparatively, in the case of neurons, each neuron received signal in the form of an action potential from the other neurons. Eventually, when the collective sum of these inputs exceeds a particular threshold value, the neuron fires by sending out signals to other neurons.

## 2.2.2 Criticality in the brain

The Brain can be described best as constantly shifting complex dynamic system and hence in this regard I draw comparison with Self Organized Complexity [1] which is one of the essential concepts to describe the emergence of complexity in natural systems.

The idea that the brain is a near-critical network is not new. A classical paper by Turing (1950) considers in passing how criticality may apply to the brain. In more recent times, "Computation at the edge of chaos" has gained considerable attention is study of complex systems (Chialvo, 2004) [15].

It has been proposed that, in order to support computation, the brain exists in near criticality (Greenfield and Lecar, 2001). Furthermore it has also been proposed that the brain self-organizes for criticality, in other words, that near critical behaviour is the consequence of the synaptic plasticity rules involved in learning (Bak, 1997).

Few of the most celebrated examples of the self-organized critical dynamics include the Sand-pile model [16], earthquakes and tectonic plate movements [17, 18] and stick-slip process [19]. Legenstein et al. [20] spoke about the edge of chaos and explored the concept of critical behaviour in bringing out optimal computational capabilities. Further critical behaviour has been seen to bring about optimal transmission [2], storage of information [21] and sensitivity to sensory stimuli. Eurich, Herrmann and Ernst [4] in their paper on "Finite-size effects of avalanche dynamics" predict the existence of critical avalanches in neuronal systems. This was further experimentally established by Beggs and Plenz [2, 22, 23].

Although in [4] critical avalanches were seen to show up only when the set of parameters were fine tuned externally to a critical transition state. Levina, Herrmann and Geisel [5] demonstrated analytically and numerically that with the more realistic assumption of dynamical synapses in a spiking neural network, the neuronal avalanches turn from an exceptional phenomenon into a typical and robust self-organized critical behaviour, provided the total resources of neurotransmitter are sufficiently large.

The importance of criticality for a neural system can be best gained by considering a feed-forward network with globally normalized connection strength $\alpha$ between two successive layers.

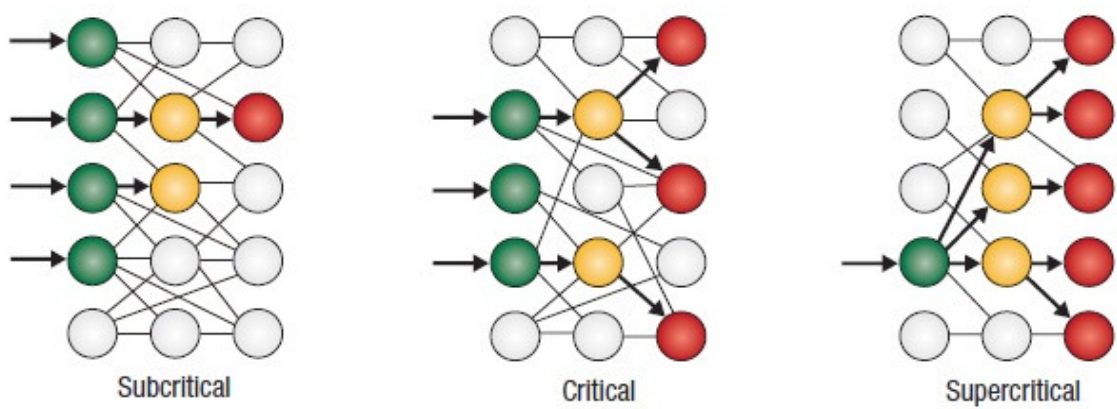

*Figure 2.2: Networks constructed with branching ratios close to one maintain, on average, the input activity (green, followed by yellow and red) and with an optimized dynamic range. There is an explosion of activity in the Supercritical networks while Subcritical networks are unable to sustain an input pattern. (Adapted from [3])*

When layer *n* sends *k* spikes, layer *n+1* will be driven to produce *l* spikes. Suppose $\alpha$ is low, this would imply that $l < k$ and the activation will fade out (sub-critical state). However if the value of $\alpha$ is high, then $l > k$, and after several steps the entire layer will spike independently of the original input. This is commonly known as the super critical state. In between these two states, there is a point of transition called the critical state, where $l = k$ on average.

Using a model of integrate fire neurons with activity dependent depressive synapses it can be shown that several dynamic regimes exist in reconciliation with parameter-independent criticality. Synaptic depression causes the mean synaptic strength to approach a critical value for certain range of values of interaction parameters, where in other dynamical behaviours are prevalent outside this range. (Figure 2.3)

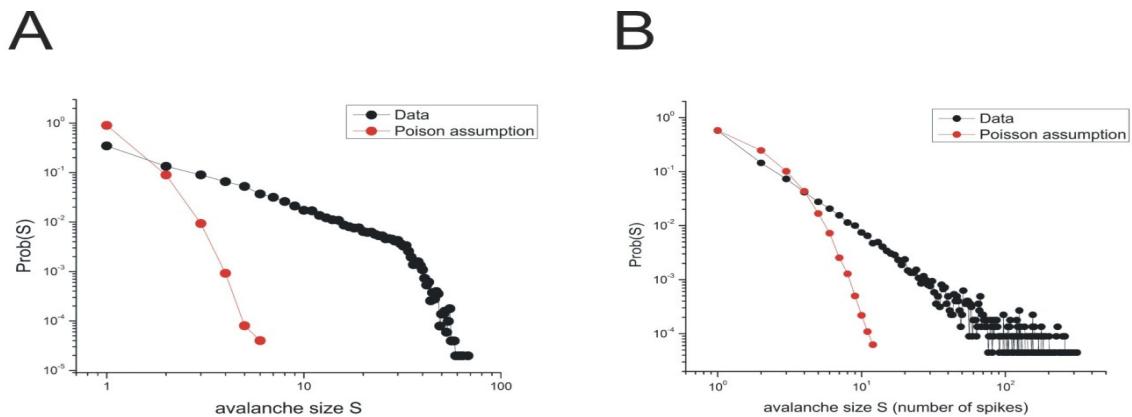

*Figure 2.3: Avalanche size distributions. A, Distribution of sizes from acute slice LFPs recorded with a 60 electrode array, plotted in log-log space. Actual data are shown in black, while the output of a Poisson model is shown in red. In the Poisson model, each electrode fires at the same rate as that seen in the actual data, but independently of all the other electrodes. Note the large difference between the two curves. The actual data follow a nearly straight line for sizes from 1- 35; after this point there is a cut off induced by the electrode array size. The straight line is indicative of a power law, suggesting that the network is operating near the critical point (unpublished data recorded by W. Chen, C. Haldeman, S. Wang, A. Tang, J.M. Beggs). B, Avalanche size distribution for spikes can be approximated by a straight line over three orders of magnitude in probability, without a sharp cut off as seen in panel A.* (**figure and data as available on Scholarpedia article by J.M. Beggs on neuronal avalanche**)

Hence it can be hypothesized that the critical state is an optimal configuration for the network. A similar idea using a branching process model (*Harris, 1989; Beggs and Plenz, 2003; Haldeman and Beggs, 2005; reviewed in Vogels et al, 2005*) can be described that captures both the power law distribution of avalanche sizes and the reproducible activity sequences observed in the data. In this model, a processing unit which is active at one time step will produce, on average, activity in $\sigma$ processing units in the next time step. The $\sigma$ number is called the *branching parameter* and can be thought of as the expected value of this ratio:

$$\sigma = \frac{\text{Descendants}}{\text{Ancestors}}$$

Where *Ancestors* is the number of processing units active at time step *t* and *Descendants* is the number of processing unit's active at time step *t + 1*. There are three general regimes for $\sigma$, as shown in the following figure (figure 2.4).

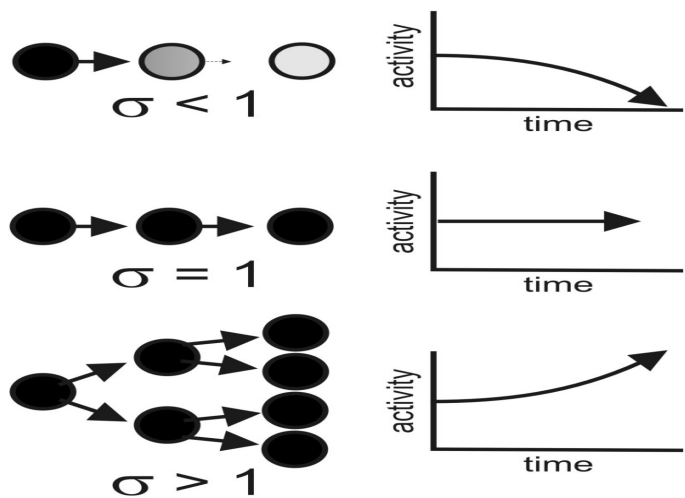

*Figure 2.4*: *The three regimes of a branching process. Top, when the branching parameter, $\sigma$, is less than unity, the system is subcritical and activity dies out over time. Middle, when the branching parameter is equal to unity, the system is critical and activity is approximately sustained. In actuality, activity will die out very slowly with a power law tail. Bottom, when the branching parameter is greater than unity, the system is supercritical and activity increases over time. . (figure and data as available on Scholarpedia article by J.M. Beggs on neuronal avalanche)*

Analysis of the information transmission shows that critical branching optimizes information throughput [2]. Furthermore the Lyapunov exponent is nearly zero at the critical point, signifying stable neural dynamics, capable of exploring all possible LFP configuration, and yet is not random [21]. Excessive spread of input excitation is caused by longer branching ratios, where by information is lost due to the intrinsic instability of the system. On the other hand rapid dampening of excitation is cause by a smaller branching ratio.

## 2.3 Neuronal avalanches

In this section we describe an important example of a self-organized critical system, especially from the point of view of this project, namely neuronal avalanches. Most importantly it was the discovery of critical avalanches in the brain that motivated this work on ageing and also renewed general interests in the field of SOC.

A neuronal avalanche can be described as a cascade of *bursts* of activity in neuronal networks whose size distribution can be approximated by a power-law (as described previously - SOC in the sand-pile model). John Beggs and Dietmar Plenz [2, 22, 23] performed the first known experiments using cultured and cortical slices, in order to demonstrate the phenomenon. Cultures were planted on multielectrode array and the local field potential signals were recorded from the 64 electrodes of the array over a long period of time (on a time scale of hours). Activity in these slices of the neo-cortex is characterized by brief bursts lasting tens of milliseconds, separated by periods of quiescence lasting several seconds.

A signal reflecting the sum of all synaptic activity within a volume of tissue [37] is commonly known as the local field potential (LFP). Although the precise origin of the signal is not yet clear, it was found that phenomenon that are unrelated to synaptic events also contribute to LFP . Furthermore, unlike spikes, the cumulative activity in the slice is better explained by LFP.

In Figure 2.1 the extraction of filtered LFP signal from the cortical slice is shown. The first filtering stage extracts LFP's from the recorded signals, which are then thresholded to obtain the binary signal on each electrode. This data is then organised in 4 ms bins. Post processing the data consists of short intervals of activity, when one or more electrodes detected LFP's above the threshold, separated by longer periods of silence. Avalanche sizes are distributed in a manner that is nearly fit by a power law. Due to the limited number of electrodes in the array, the power law begins to bend downward in a cut-off at the size of the multi-electrode array. But for larger electrode arrays, the power law is seen to extend much further. The exponent of -3/2 is seen to characterise both acute cortical slices and cultures.

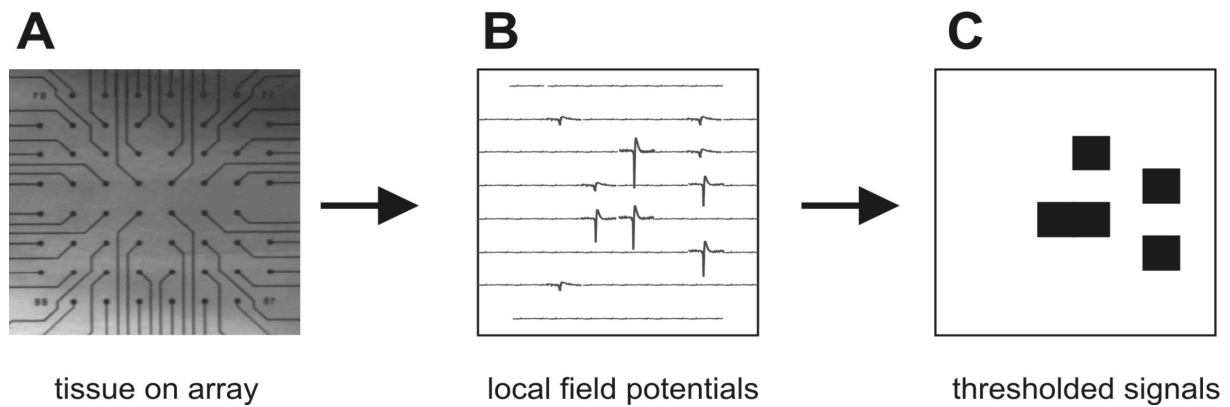

**Figure 2.1**: *Extraction of LFP's from a multielectrode array (Image taken from Levina [41])*

Interestingly, models that explicitly predicted avalanches of neural activity include the work of Hertz and Hopfield (1995) [18] which connect the reverberations in a neural network to the power law distribution of earthquake sizes. They considered leaky integrate-and-fire neurons with a constant input current on a lattice and were mainly interested in synchronization of such a network in the non-leaky case.

First example of a globally coupled system showing criticality was presented by Eurich, Hermann and Ernst (2002) [4].They predicted that the avalanche size distribution from a network of globally coupled nonlinear threshold elements should have an exponent of $\alpha = 1.5$. Remarkably, this exponent turned out to match that reported experimentally (Beggs and Plenz, 2003) [2]. They investigated a simple model of a fully connected neural network of non-leaky integrate fire neurons. The results predicted the critical exponent as well as other different dynamical phenomena. In the works of Arcangelis et. al [42,43] self organised criticality is obtained from a different perspective. The authors considered the network on a grid using anti-hebbian learning rule. It was used to explain the power law distribution of the EEG spectrum.

More interestingly in recent times, Levina et al. (2007 & 2009) [5, 25] analytically proved that Depressive synapses caused a broad and stable critical regime in Hopfield networks independent of the parameter settings.

# Chapter 3

# Approach: Does learning disrupt criticality?

This chapter explores the main approach and methods used to establish our primary hypothesis "Does prolonged learning disrupt criticality" (as introduced in chapter 1) in greater depth. Initially the model of Levina et al. [5, 25] from 2007 and 2009 is discussed with the description of the concept of dynamical synapses and associative memory in order to discuss the primary approach taken in this project and also to lay the foundations for modified versions of this model. This is followed by the variant of the original Hopfield model with modified Hebb's rule used and the limitations of the standard model thus taken care of. Following these, descriptions of the modification of the original model including model with homeostatic regulation is presented, along with mathematical justification and discussion of the method undertaken. Finally we introduce two important evaluation parameters of simulation results.

## 3.1 Dynamic synapses

In this section we briefly describe the biological background and model necessary to understand *dynamic synapses,* a special type of short term synaptic plasticity. Thomson and Deuchars [50] were the very first to observe them and noted that transmission across neocortical synapses depends on the frequency of pre-synaptic activity, although the nature of this dynamic transmission is seen to vary between different classes of neurons.

The phenomenological model of neocortical synapses provides an abstract model of neuronal dynamics by computing the post-synaptic current generated by a population of neurons with specific firing rates. The model assumes that a synapse is characterized by a fixed amount of "*resources*". Each pre-synaptic action potential (arriving at $t_{sp}$) activates a particular fraction of resources, which then quickly inactivate within a time constant of few milliseconds ($\tau_{in}$) and then recovers with a time constant of about 1 second ($\tau_{rec}$).

Mathematically the kinetic equations corresponding to this model is as follows:

$$\frac{dx}{dt} = \frac{z}{\tau_{rec}} - U_{SE} x \delta(t - t_{sp})$$

$$\frac{dy}{dt} = -\frac{y}{\tau_{in}} + U_{SE} x \delta(t - t_{sp})$$

$$\frac{dz}{dt} = \frac{y}{\tau_{in}} - \frac{z}{\tau_{rec}} \quad (3.1)$$

Here *x, y* and *z* are the fractions of resources in the recovered, active and the inactive states respectively. The post-synaptic current is considered to be proportional to the fraction of resources in the active state. Furthermore Tsodyks and Markram [26] showed that for individual synapses, this model reproduces the post-synaptic resources generated by any pre-synaptic spike train $t_{sp}$ for inter-pyramidal synapses [49] in the layer V. It is also known that a higher frequency of firing leads to the smaller effect of each spike and hence the term depressing may be attributed to these synapses. More specifically depressing synapses are one of the mechanisms that connect the rate code with the precise timing code.

## 3.2 Modelling associative memory with dynamical synapses

The primary idea in order to explore our hypothesis was to consider a neural memory model that allowed associative recall combined with the display of self-organised criticality in the network. As a first step we implemented learning in a neural network with dynamic synapses based on the model proposed by Levina et al. 2007.

Previously the strength of interactions among the neural units by synaptic connections has been identified as critical parameter to achieve SOC [4]. However in the case of biologically realistic neural systems the connection strengths are never static but rather depend on the relative timing of the pulses of neuronal activity. Hence in this regard it is more justified to consider dynamical synapses as shown by [5]; where in it was demonstrated that dynamical synapses cause typical and robust self-organised critical behaviour in the neural network.

Interestingly electrophysiological recordings have shown that there is a possibility of change in the amplitude of the postsynaptic response of biological neurons to incoming spikes, under repeated stimulation [48]. Moreover the weakening of the synaptic strength has proven to be a characteristic property of synaptic connections between pyramidal neurons in different areas of the brain. Theoretical studies in this field have shown that the dynamical properties of synapses can have an influence on the transmission properties of single neurons as well as the network level dynamics.

Having discussed these concepts, we now introduce the basic neural model with activity-dependent depressive synapses that is shown to exhibit self-organised criticality in a parameter independent manner. As mentioned previously the Levina et al. model served as the foundation for our modification to combine Hebbian learning with parameter independent criticality. In order to incorporate the behaviour for real neurons and evaluate the influence of forgetting of memories we introduce homeostatic regulation of the average synaptic strength between two neurons.

### 3.2.1 A neural model for criticality

When a specific external input signal is input into the brain, a particular location of cortex responds with an increase in the membrane potential of response neurons. When the membrane potential of a neuron exceeds a specific threshold value, the neuron releases signals in the form of action potentials and then returns back to the rest state the neuron had fired from. This signal is transferred by the synapses to the other neurons, which has an excitatory or inhibitory influence on the membrane potential of the receiving cells according to whether the synapses are excitatory or inhibitory, respectively. The resulting potential, if it also exceeds the threshold, leads to the next firing step, and so on, resulting in a neural avalanche. It will then cause a response in some other areas of the cortex. Based on this phenomenon the following model is conceived.

## 3.2.2 Network with depressive and facilitative synapses

Our approach is based on the model of dynamical synapses, which was shown by Tsodyks and Markram (1996) to reliably reproduce the synaptic responses between pyramidal neurons. A fully connected network of $N$ simple integrate-fire neurons that interact by exchanging short pulses of activity is considered. The spikes or action potentials generated are transformed into chemical signals that are transmitted between two neurons across the synaptic cleft. Each neurons is characterized by a membrane potential $0 < h_i(t) \leq \theta, i = 1,....,N$, where $\theta$ is a threshold unit.

A random process $\varsigma \in \{1,...,N\}$ serves as the external input to the neuron which with the rate $\tau_s$ selects a neuron $\varsigma(t) = i$ whose membrane potential $h_i$ is advanced by an amount $I^{ext}$. Until the threshold is reached it neuron integrates inputs signals. When the membrane potential $h_i(t) > \theta$ at time $t_{sp}^i$ the neuron $i$ emits a spike which is then delivered to all postsynaptic neurons at a fixed delay $\tau_d \ll \tau_s$. This is then reset by subtracting the threshold $\theta$: $h_i(t_{sp}^+) = h_i(t_{sp}) - \theta$. The amount of neuro-transmitter emitted at a synaptic terminal is dependent on the relative timing of the arriving pulses exclusively. This is denoted by the term $J_{ij}$ and the fraction, and the particular fraction that is available for signalling at any given moment by $u_{ij} \in [0,1]$. The product $u_{ij}J_{ij}$ represents the interaction strength. An avalanche of neuronal activity is triggered as the neuron reaches the threshold value, upon receiving input from the external source. An avalanche is defined here as described in Section 2.3.

Mathematically this dynamics of the membrane potential can be expressed as:

$$\vec{h}_i = \delta_{i\varsigma(t)} I^{ext} + \frac{1}{N} \sum_{j=1}^{N} u_{ij}(t_{sp}) J_{ij}(t_{sp}) \delta(t - t_{sp}^j - \tau_d) \quad (3.2)$$

Where, $\delta_{i\varsigma(t)}$ is the Kronecker symbol and,

$\delta_{i\varsigma} = 1$; If $i = \varsigma$ and $\delta_{i\varsigma} = 0$ otherwise. Further via the Dirac delta function,

$$(\delta(t) = 0 \text{ If } t \neq 0 \text{ and } \int \delta(t)dt = 1)$$

$\tau_s$ : represents the rate of external input.

The Dynamics of $J_{ij}$ can be described as follows:

$$\vec{J}_{i,j} = \frac{1}{\tau_J}(\frac{\alpha}{u_o} - J_{i,j}) - u_{i,j}J_{i,j}\delta(t - t_{sp}^j) \tag{3.3}$$

Basically the synapses are subject to a loss of neuro-transmitter when they transmitted a as spike, reducing their efficacy $J$ by a fraction $u_{ij}$.

The dynamics of $J_{ij}$ can essentially be seen to be composed of two parts:

- A decrease in the strength of the synapse occurs upon activation; as there is a depletion of the resources of the synaptic transmitters.
- While the neuron is silent, the synapses slowly recover. Thus in between spikes, the resources recover and $J_{ij}$ approaches the resting value of $\frac{\alpha}{u_o}$.

For experimental purposes and without the loss of generality in the simulations we use the threshold value $\theta = 1$.

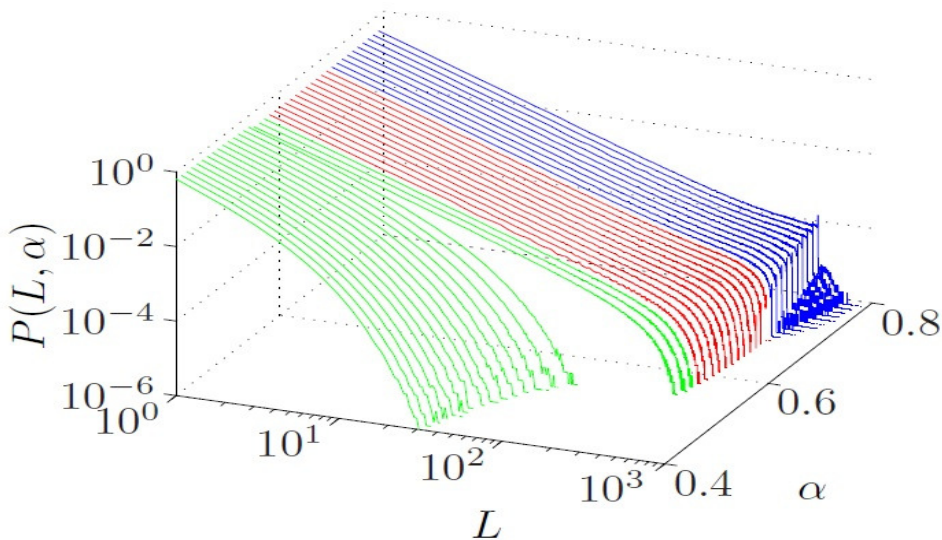

*Figure 3.1*: *Distribution of avalanche sizes for different coupling strengths $\alpha$ is plotted. (Continued on next page).*

*It is seen that at $\alpha < 0.54$ the distribution is subcritical (coded in green). The range of connectivity near $\alpha = 0.65$ appears to be critical (Red). For $\alpha > 0.69$ the distribution is in the supercritical zone (blue). These results were obtained for a small sized network of 200 neurons with $u_o = 0.1$. Here L is the size of avalanche and $P(L,\alpha)$ the probability distribution. (figure reproduced from [25]).*

We were able to exactly reproduce the same results obtained by Levina et. al [25], namely, it is observed that for the basic model with depressive and facilitative synapses, it is seen that several dynamical regimes coexist with parameter independent criticality. Using Fig 3.1 obtained from the simulation results of the above model, it can be observed that the synaptic depression causes the mean synaptic strengths to approach a critical value for certain range of interaction parameters, while outside of this range other dynamical behaviours exist. A clear power law distribution is visible with the three regimes perfectly distinguishable (Figure 3.2).

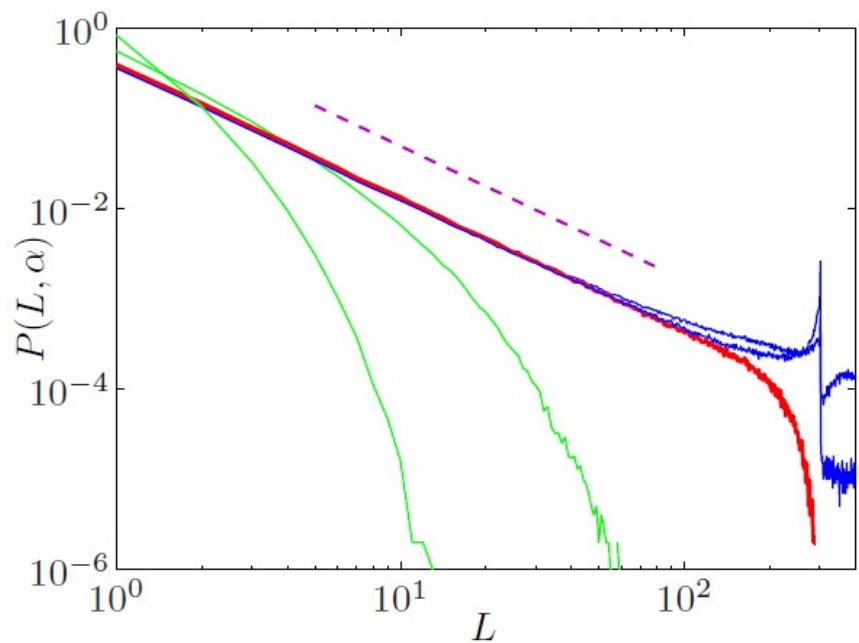

*Figure 3.2: A depiction of the three dynamical regimes: sub-critical (green), critical (red), super-critical (blue). The dotted line shows the power law distribution with a measured slope or exponent of 1.5. (Image adopted from [25])*

Having explored self-organized criticality with dynamical synapses, we modified the initial model in order to combine Hebbian learning with the observed dynamics of the prior model. We considered the modified Hebb's rule (Equation 1.15) for optimal learning of associative memory with dynamic synapses.

## 3.2.3 Combining Hebbian learning with SOC

Using the same network of *N* simple integrate-fire neurons, we stored a set of *p* biased-correlated [32] patterns $\xi_i^\mu$, where $\mu = 1,...,P$ and $i=1,....., N$. The dynamics of the emitted neuro-transmitter $J_{ij}$ are modified such that, the synapses are subject to a loss of transmitter when they transmitted a spike reducing their capability $J_{ij}$ by a fraction $u_{ij}$, however unlike the previous case the synapses now recover toward the maximum synaptic strength or coupling between two neurons $W_{i,j}$, as obtained from the modified Hebbian rule and Hopfield statistics, with a slow time scale of $\tau_J$.

**The dynamics can be represented as follows**:

From Hebbian rule (1.15) the correlation matrix $W_{ij} = \dfrac{1}{Nf(1-f)} \sum_{\mu=1}^{P} (\xi_i^\mu - f)(\xi_j^\mu - f)$;

Where, $f = \dfrac{\sum \xi_i^\mu}{N}$ and $\xi_i^\mu \in \{0,1\}$

$$\vec{J}_{i,j} = \dfrac{1}{\tau_J}(W_{i,j} - J_{i,j}) - u_{i,j} J_{i,j} \delta(t - t_{sp}^j) \qquad (3.4)$$

Furthermore Markram and Tsodyks (1996) provided experimental evidence that the value of $u_{ij}$ is also subject to an activity dependent dynamics. Therefore this can be represented as,

$$\vec{u}_{i,j} = \dfrac{1}{\tau_J}(u_0 - u_{i,j}) + (1 - u_{i,j}) u_0 \delta(t - t_{sp}^i) \qquad (3.5)$$

With each firing event there is an increase in the probability of other neurons being activated such that a number of neurons may join the externally triggered activity and create a neural avalanche. Synaptic depression becomes dominant when the avalanches are large and the neurons fire frequently and causes a reduced activity and hence smaller avalanches. Contrary to this sparse firing events lead to almost full recovery of the synapses, and facilitation of synapses becomes essential [25]. However the distribution of the number of neurons participating in an avalanche still depends on the value of $\alpha$. From figure 3.1 qualitative changes can be seen to occur at the critical value of $\alpha_c = 0.543$. For $\alpha < \alpha_c$ the subcritical region is characterised by a negligible number of avalanches extending to the size of the system. Non-monotonic distribution is observable for large avalanches when $\alpha >> \alpha_c$. Moreover it is important to note that in between two spikes of neuron $j$ only the relaxation dynamics affects the variable $u_{ij}$

## 3.3 Homeostatic regulation: Realistic leaky neurons

It is important to note here that there are numerous complex mechanisms going on at a synapse. Most interestingly synaptic adaptation outside of Hebbian learning in the form of Homeostatic plasticity [29] is thought to balance Hebbian plasticity by modulating the activity of the synapse or the properties of ion channels. During the process of learning and development, a refinement of neural circuitry occurs, partly due to the changes in the number and strength of synapses. Turrigiano [28, 30] in her 1999 paper on "*Homeostatic plasticity in neuronal networks: the more things change, the more they stay the same*" noted that though Hebbian mechanisms are crucial for the selective modification of neuronal circuitry, they may not be the sufficient. Furthermore various forms of homeostatic plasticity that stabilize the properties of neural circuits have been identified.

In real neuron during the process of synaptogenesis, axons and dendrites undergo a process of dynamic extension and retraction that allow many pre and postsynaptic elements to come in contact with each other to form 'trial' synapses [28]. However there needs to be a regulation of this growth and retraction by activity at the synapse; such that further strengthening of the synapse halts growth and stabilizes the pre and post synaptic structures. At the same time weakening of the synapse allows the pre and post synaptic elements to continue to grow.

Turrigiano and Nelson [29] in their discussions of homeostatic plasticity focus on mechanisms for maintaining a stable firing rate. It was noted that chronologically reducing inhibition in cortical networks initially raises firing rates, but eventually the firing rates return to control levels [47]. Although these experiments suggested that neurons in the network have some specific point of activity that is dynamically maintained , the exact feature of activity that is conserved is still not completely clear, and mean firing rates, mean calcium concentrations or some more subtle measure of neuronal activity can be possible candidates. However it is established that different aspects of network function could be modulated to maintain the set point of activity. These include the strength of excitatory and inhibitory connections and also the intrinsic excitability of individual neurons (Figure 3.3).

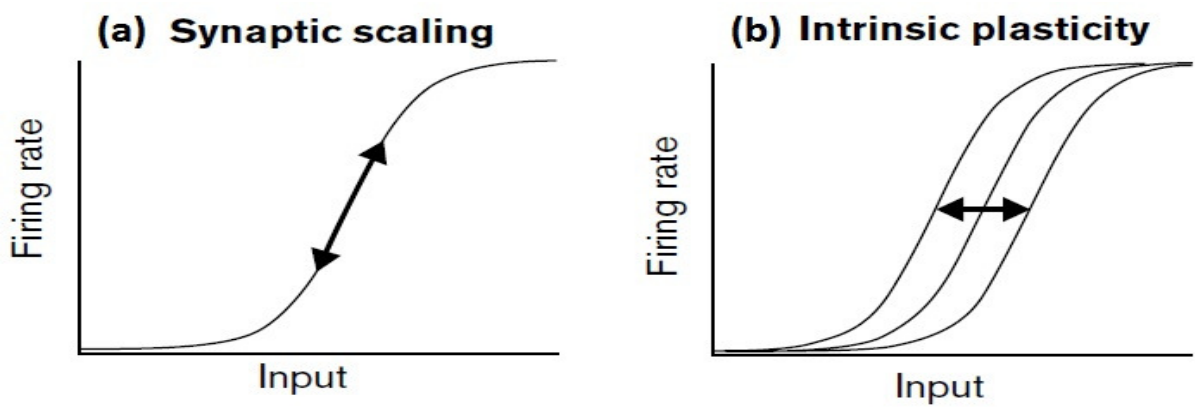

**Figure 3.3**: **(a)** *By scaling the strength of all of the neurons inputs up or down, the neuron's property can be shifted up or down its input/output curve. This determines how fast the neuron fires for a given amount of synaptic drive.* **(b)** *Contrastingly the regulation of intrinsic conductance (plasticity) can modify the input/output curve of the neuron by shifting it right (it fires more for a given synaptic drive) or left (it fires less for a given level of synaptic drive).It can also modify the slope of this curve.* (Figure with modifications from [29]).

However it is important to note that, when we talk about criticality in the network, the concept of "firing rate" is not very clear. This is mainly because of the fact that the dynamics of the system is characterized by very irregular firing of the neurons. As such the mean firing rate is not an optimum characterization of the neural activity.

We believe that these same homeostatic mechanisms also ensure criticality. This means that the average strength of the synapses should be at the critical level for leaky (i.e. real) synapses. This is necessary in addition to the depression and facilitation of the neurons to maintain criticality, for a biologically more realistic scenario.

### 3.3.1 Homeostatic regulation of synaptic weights

Keeping these observations and research findings on Homeostasis in mind, we make further changes to the model described above, and introduce a synaptic adaptation outside of the Hebbian learning such that after the occurrence of a spike the amount of neuro-transmitter $J_{ij}$ is reduced and then slowly approaches the maximum synaptic strength $W_{i,j}$ in between the spikes, in addition the average strength of the synapses also move towards the critical level, controlled by the arrival of a postsynaptic spike. This summarizes neuronal self-regulatory mechanisms. In this way the probability of the neuron to stay near the threshold is conserved and neuronal avalanches are triggered in a similar way as in the non-leaky case. Although we haven't tested this model extensively with regards to the retrieval quality of the stored patterns (*m*), the use of modified updating of the synaptic weights seems to conserve the criticality of the system.

The modified dynamics can be represented as follows:

$$\vec{h}_i = \delta_{i\varsigma(t)} I^{ext} + \frac{1}{N} \sum_{j=1}^{N} u_{ij}(t_{sp}) J_{ij}(t_{sp}) \delta(t - t_{sp}^j - \tau_d)$$

$$\vec{J}_{i,j} = \frac{1}{\tau_J}(W_{i,j} - J_{i,j}) - u_{i,j} J_{i,j} \delta(t - t_{sp}^j)$$

$$\vec{W}_{i,j} = \vec{W}_{j,i} = \sum_{j=1}^{N} k_o.(\alpha - W_{i,j}).\delta(t - t_{sp}^j) \quad , \tag{3.6}$$

Where $k_o = 0.1$ is a constant, set experimentally.

$$\vec{u}_{i,j} = \frac{1}{\tau_J}(u_0 - u_{i,j}) + (1 - u_{i,j}) u_0 \delta(t - t_{sp}^i)$$

Dynamics of the membrane potential remains the same.

For simulation purposes, these models are used with varying network sizes from N = 150, 200, 300 and varying pattern numbers from $p$ = 1, 5, 10, 25 (such that $\frac{P}{N}$ is not $\geq 0.1$). Specifically the single fixed pattern scenario and orthogonal patterns are of high interest. (details in chapter 4). Moreover different pattern types from random, correlated hierarchical to orthogonal patterns were tested and the effect of prolonged learning of these on the criticality of the network was observed (Section 4.1).

As mentioned in Chapter 1, the primary parameters for evaluation of the simulation results were *mean-squared deviation from power-law* ($\Delta\gamma$) and the measure of the *retrieval quality of patterns* (*m*) using mean field techniques. In Section 3.2 we introduce these two parameters in greater detail and explore their evaluation, values and general trend for the original self-organized critical model and then standard stochastic Hopfield network. In the next chapter the results of the simulation are discussed in depth with further evaluation of the two parameters in the context of the modified models.

## 3.4 Model evaluation parameters

The simulation results are evaluated primarily based on the estimation of power-law parameters (By observing the probability distribution graph on a log-log scale. E.g. Figure 3.1.) and the comparison to the parameter values of the original SOC model [5] along with evaluation of the degree of overlap or the quality of retrieval of the stored patterns. Thus an evaluation of the mean squared deviation from power-law along with the retrieval quality gives us an understanding of the change in the dynamics of the system during the learning and recall of the stored patterns and its effect on the criticality of the network. Here we present a mathematical explanation of these two important evaluation metrics with respect to the standard SOC model and associative memory with dynamic synapses.

### 3.4.1 Mean squared deviation from power-law

A quantity *x* obeys a power law if it is drawn from a probability distribution

$$P(x) \propto x^{-\gamma} \qquad (3.7)$$

Where $\gamma$ is a parameter of the distribution known as the exponent or scaling parameter. Although in general it is difficult to say with absolute surety if the observation was drawn from a power-law distribution. However for certain intervals of the distribution these observations may be consistent with equation 3.6. Hence as a first step towards verification of the results is given the simulation results, to specify the parameter $\gamma$ and determine whether the results truly follow a power-law distribution with this exponent. Taking logarithms on both sides of equation 3.6 we see the power-law distribution obeys $\ln p(x) = \gamma \ln x + const$, implying that it follows a straight-line on a double logarithmic plot. We use least squares linear regression technique on the logarithm of the histogram of $x$ in order to estimate the power-law exponent $\gamma$. In order to deal with power-laws with an exponential cut-off, the exponent $\gamma$ is calculated using a maximum-likelihood estimation technique (**Appendix A.2**).

The mean squared deviation $\triangle \gamma$ is calculated between the interval $l = 2$ to $N=2$ due to the finite size of the avalanche distribution and the maximum avalanche size being fixed at the size of the network.

$$\triangle \gamma = \frac{1}{N/2} \sum_{l=2}^{N/2} [\ln P(l) - a(\ln l + b)]^2 \qquad (3.8)$$

Hence solving this for $\frac{\partial (\triangle \gamma)}{\partial a} = \frac{\partial (\triangle \gamma)}{\partial b} = 0$ we obtain the slope and the least- squares error.

Where $l$ is the size of the avalanche, $P(l)$ is the probability distribution depicting the network dynamics and $a$, is the slope of the power-law. However it is important to note that criticality is defined using $\triangle \gamma$ for finite networks, where as only infinite or biologically realistic networks would show exact criticality.

## 3.4.2 Quality of retrieval of stored patterns

In the analysis of the associative memory model, we are primarily interested in the retrieval quality of the stored patterns (i.e. degree of overlap) in conjunction with the above mentioned mean squared deviation from the power law statistics. An important question while looking at the effect of Hebbian learning on criticality is to test whether the fixed points of the system dynamics $s_i(t+1) = \sigma(h_i(t) - \theta)$; stay close to the stored patterns as the network operates with depressive and facilitative synapses.

The closeness to a certain pattern $\xi^\mu$ is characterized by an overlap

$$m^\mu = \frac{1}{Nf(1-f)} \sum_{i=1}^{N} \eta_i s_i(t), \text{ Where } \eta_i = \xi_j^\mu - f \tag{3.9}$$

Which can be seen to vary in the interval $-1 < m^\mu \leq 1$. Using the free-energy and saddle-point equations from Amit et. al [32] it can be seen that,

$$m^\mu = \left\langle\!\!\left\langle (\xi^\mu - f) \tanh\left[\beta \sum_\mu m^\mu ((\xi^\mu - f)\right]\right\rangle\!\!\right\rangle \tag{3.10}$$

Moreover since,

$$\left\langle\!\!\left\langle (\xi^\mu - f)(\xi^\nu - f) \right\rangle\!\!\right\rangle = 0$$

It follows that there are always retrieval state solutions, $m^\mu = m\delta^{\mu,\nu}$ for which,

$$m = \frac{1}{2}(1-f^2)\left[\tanh \beta m(1-f) + \tanh \beta m(1+f)\right] \tag{3.11}$$

Therefore in the case of a perfect overlap of the retrieved pattern with that of the $\nu^{th}$ stored pattern,

$$\bar{m} = \frac{1}{N} \sum_i (\xi_i^\mu S_i) = 1 \tag{3.12}$$

As mentioned previously, the stored patterns act as basins of attractions, which can be seen to gradually shrink as more and more patterns are added. As a result of this we study the change or effect of $m$ (3.9) over specific intervals of time as the system dynamics changes. In [31] for attractor networks with short-time synaptic dynamics they found the existence of essentially three types of stationary states, namely pattern retrieval with $m \approx 1$, failure with $m \approx 1$ and the oscillation states, in which groups of neurons from patterns, which have certain overlap with each other, are alternately active for short period of time. Both perfect retrieval and the oscillatory states are cleared observed in the simulation results of our neural model and in this regard their dynamics is studied to gauge the effect of this on criticality of the system.

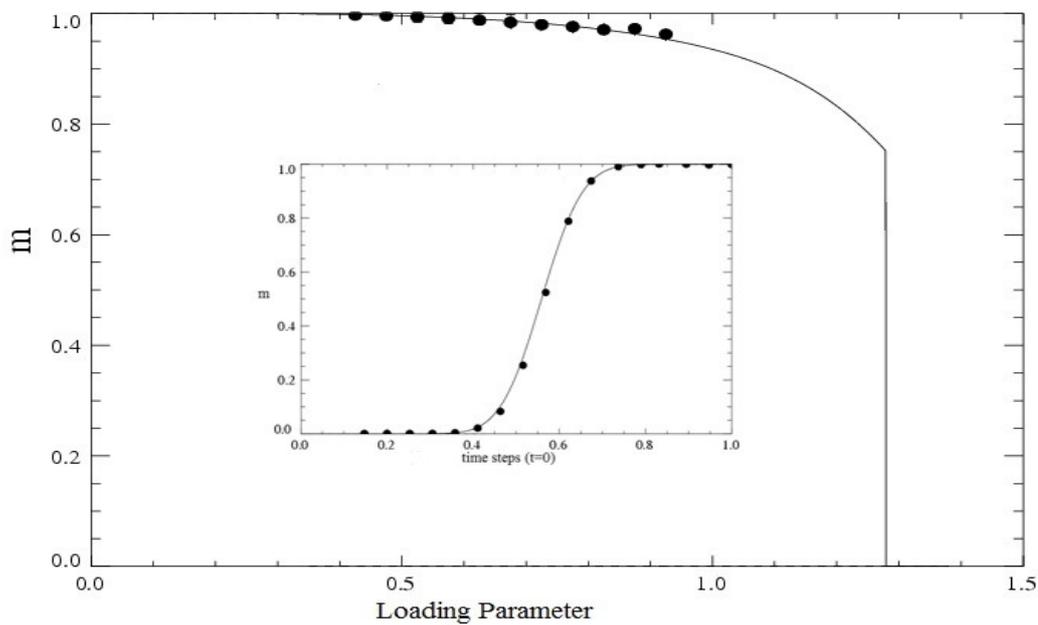

*Figure 3.4: Overlaps of memory states as function of loadings in the network. The inset shows the overlaps of retrieved patterns as functions of initial overlaps m(t=0) for loading with $\alpha^* = 0.1$. (Reproduced from [31]).*

In order to test the above mentioned overlap equations, a standard Hopfield network with finite biased correlated patterns is simulated and the retrieval quality with respect to different loading parameters ($\alpha^* = \frac{P}{N}$) as well as function of initial overlap at time step t=0 is observed. (Figure 3.4)

# Chapter 4

# Results & Evaluation

In chapters 2 & 3 the necessary background along with the approach taken to establish our original hypothesis and the neural models thus used, were introduced and discussed. We also described the primary evaluation parameters used in order to make our claim as being plausible. In this chapter, we demonstrate the experimental procedures and simulations which were followed in order to evaluate each of the 3 neuronal models (as introduced in Section 3.2.2, Section 3.2.3 and Section 3.3.1)

## 4.1 Model I: Network with dynamic synapses

The distribution of the avalanche size or different values of the parameter $\alpha$ confirm the fact that the mean synaptic strengths of the neurons approach a critical value as a result of synaptic depression. From the dynamics of the model it is known that a super-threshold activity is communicated along neural connections of strength proportional to $J_{i,j}$ to the other neurons and in turn may cause them to fire. It is in this way that an avalanche of neural activity of size $L \geq 1$ is triggered. Although traditionally the size of the avalanche is defined as the number of participating neurons, in this simulation the value of $L = N + 0.1N$ so that we allow the avalanches to continue with diminishing activity for some time in order to observe the dynamics of the super-critical region.

In Figure.4.1 it is observed that there is a sharp transition occurring at the $\alpha = 0.54$ from the sub-critical state in to a critical state, as indicated by the straight line slope of the log-log graph, clearly indicating a power-law distribution. Furthermore in order to obtain the characteristics of the observed power-law, we observe the parameters of the fitted power-law.

As highlighted in Section 3.2, for the purpose of obtaining the width of the critical region and comparison of the effect on criticality of the different models, the mean squared deviation ($\Delta\gamma$) of the avalanche size distribution from the exact power law for different values of $\alpha$ is plotted. It is to be noted that although we run the experiments keeping the network size fixed at 200 or 500 neurons, the general trend remains the same for larger networks with just the width of the critical region increasing with the system remaining critical for a substantial part of the parameter space.

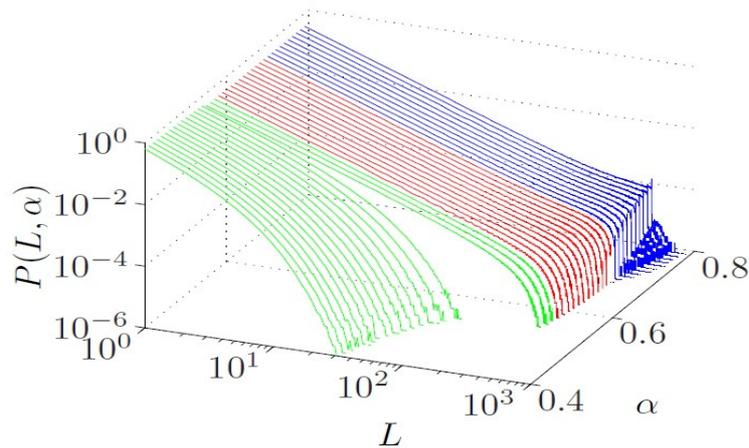

*Figure 4.1*: *The observed distribution of avalanches for different values of $\alpha$. It is seen that at $\alpha < 0.54$ the distribution is subcritical. The range of connectivity near $\alpha = 0.65$ appears to be critical. For $\alpha > 0.69$ the distribution is in the supercritical zone. (Both x and y-axis are on log scale)*

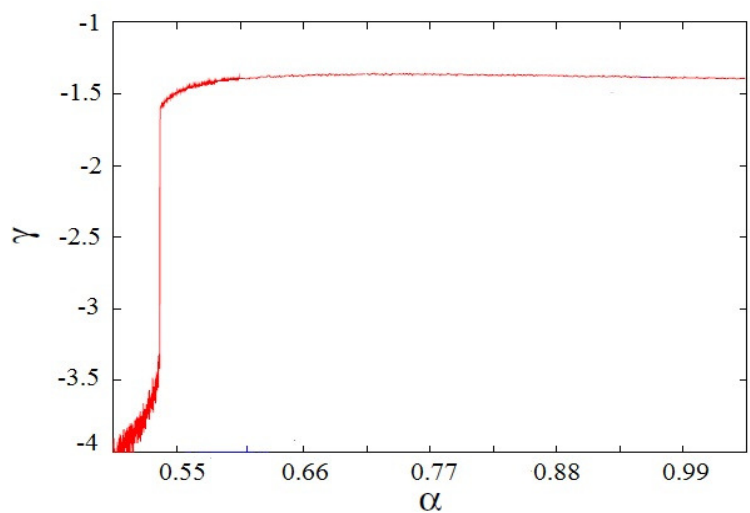

*Figure 4.2*: *(In the previous page) this shows the best matching power-law exponent fitted according to the avalanche sizes greater than 2 and smaller than N/2. The synaptic efficacy α varies from 0.33 to 0.99 with a step size of 0.01. (Figure adopted from [5])*

## 4.2   Model II: Combining Hebbian learning with SOC

Here we consider a neural memory model of N-single integrate and fire neurons that allows or associative recall, and whose dynamics is as specified in **Section 3.2.3**. The primary goal was to use the simulations in order to evaluate and establish our original hypothesis that criticality of the system is affected due to prolonged learning in the network. Furthermore we discuss the ability of certain pattern types to affect the criticality of the system which could possibly shed more light on the type of stable memory encoding inside the brain.

For all the simulations carried out with this model, a number of patterns (subset of neurons) are chosen and the same sets of patterns are shown the entire learning procedure. As discussed previously power-law fitting error and the retrieval quality of the stored patterns are used as the primary evaluation parameters. In order to qualify the difference between the different cases we define a threshold describing a "very good fit" and compared the parameter regions which deliver a critical distribution with at least this qualify. However the result is independent of a choice of threshold value as long as it is not too large. Further it is observed that the critical region increases with an increase in the size of the network, suggesting that in the thermodynamical limit the network is critical for a substantial fraction of the connectivity parameters.

The Different pattern sets used are as follows:

- Single distinct pattern of 50 neurons.
- 20 distinct patterns of 50 neurons each.
- 20 randomly chosen patterns of 50 neurons each.
- 5 Orthogonal patterns of 50 neurons each.

In all the four cases the simulations were run with the following fixed parameters (keeping space-time complexity of the experiments in mind):

Number of neurons is 300, $u_0$=0.1, Maximum avalanche is 320 and the value of $\alpha$ varies from 0.33 to 0.99, incremented in steps of 0.01.

## 4.2.1 Single fixed pattern case

In Fig 4.3 we observe the distribution of the avalanche size for different values of the parameter $\alpha$ and characteristically a clear power-law is observed with a clear transition from sub-critical to critical state, however with a shift of the critical region to higher $\alpha$ values. Interestingly a sudden loss of dynamics in the form of a gap in the critical region is observed for a small duration of higher $\alpha$ values.

For the purpose of comparison with the results of the previous model the mean squared deviation of the avalanche size distribution from the power law for different $\alpha$ and fixed N are also plotted.

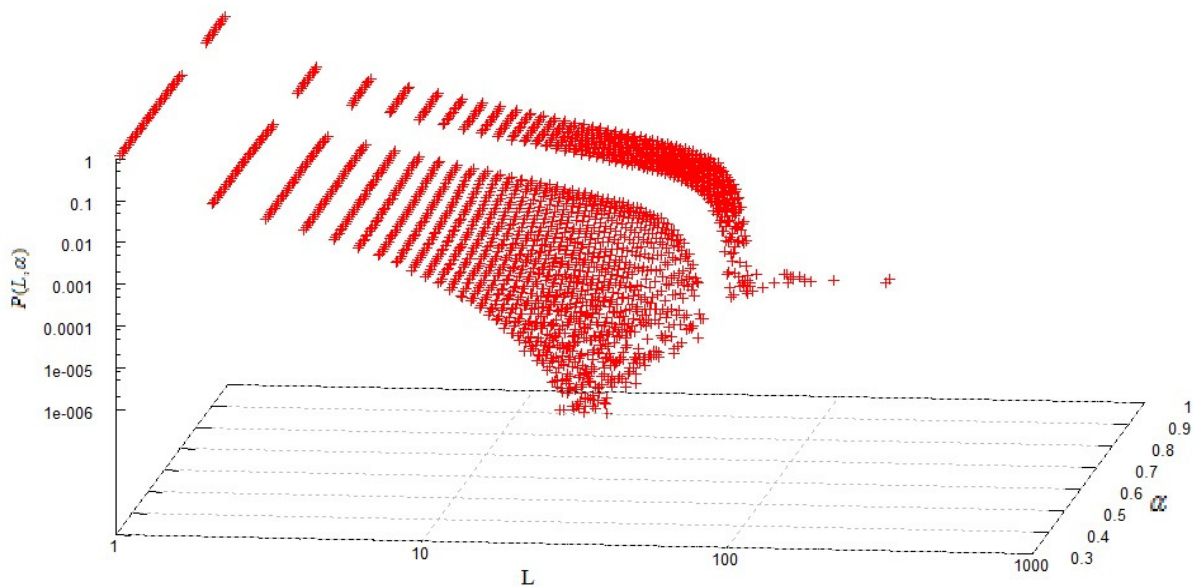

*Figure 4.3*: *Observed dynamics of the avalanches in the system of associative memory with dynamical synapses for a single distinct pattern. A power law slope is clearly visible along with an interesting gap for small range of high $\alpha$ values.*

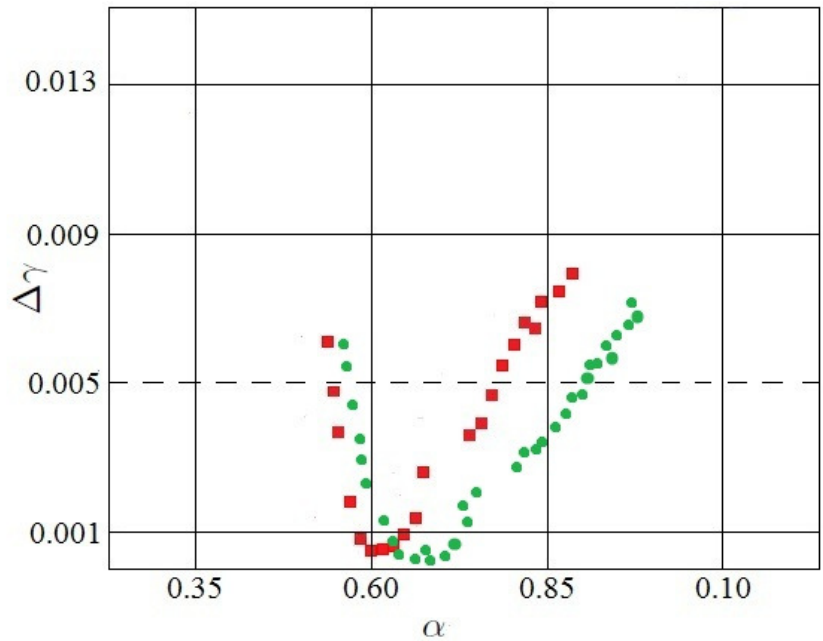

*Figure 4.4*: *Here the mean squared deviation from the best matching power-law is plotted in dependence of $\alpha$. The red squares, green circles stand for networks with depressive and facilitative synapses and system sizes N=300,500, respectively.*

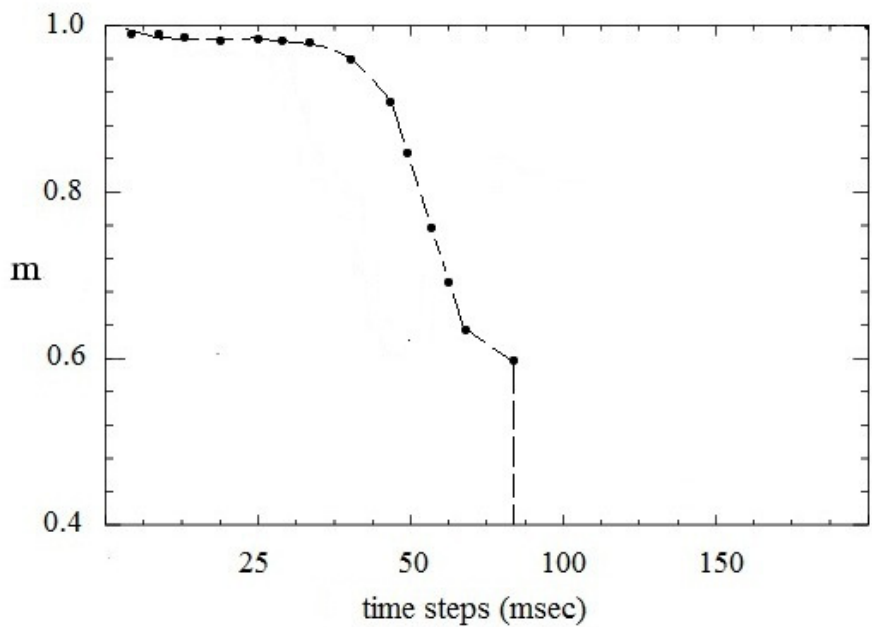

*Figure 4.5*: *Change in retrieval quality dynamics during at different time steps for a single pattern case. A step like formation at relatively high retrieval quality indicates the presence of possible overlaps between neurons within the pattern.*

From Fig. 4.4 we can clearly observe a shift in the criticality. However a large portion of the values still lie well below the threshold of 0.005 suggesting power-law distribution. Although criticality is maintained, the point of criticality is now occurs earlier than the previous case i.e. the range of connectivity near $\alpha = 0.62$. Furthermore comparing this to the mean-squared deviation graph of the original model certain distinct changes are visible, however it is still difficult to completely account for the sudden absence of criticality for certain region of the space and then re-appearance of the critical zone. Nevertheless it is highly likely that sudden destruction of power-law or the gap in the graph is created due to overlaps between neurons (Fig 4.5) within the single pattern, affecting the synaptic strength between the neurons and in turn momentarily destroying the criticality. However it is clear that learning of a single fixed pattern does not affect the criticality in the network

.

## 4.2.2 Multiple pattern case:

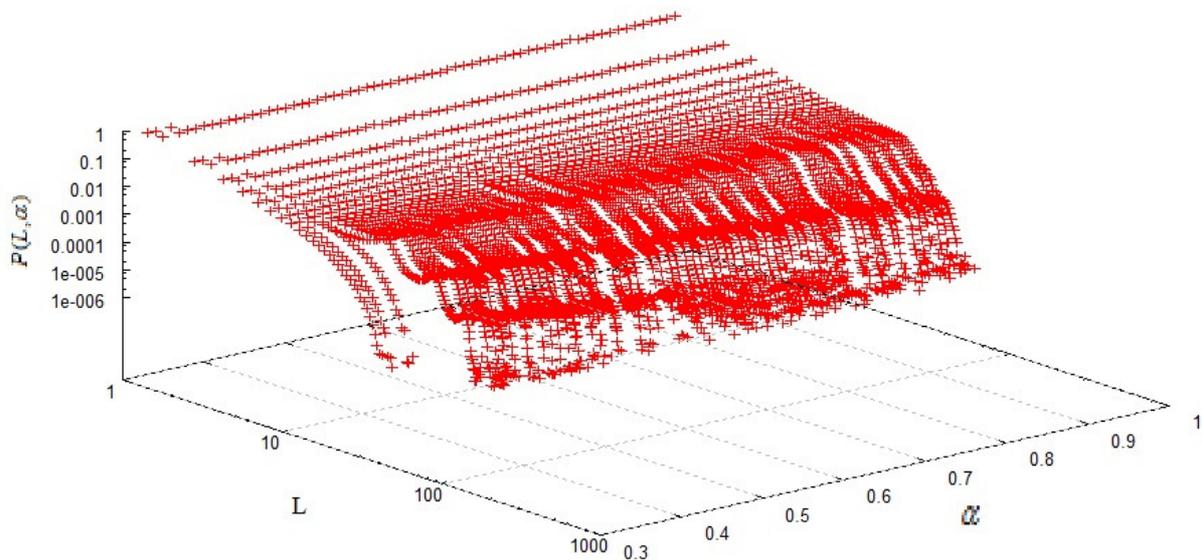

*Figure 4.6*: *The observed avalanche distribution of the network for the random multiple patterns. A small section corresponding to the subcritical regime is observed followed by an immediate transition into a highly structured regime. This prevents the onset of a critical regime. 20 random patterns are used for a network with 300 neurons.*

Initially the network dynamics is evaluated using both the fixed and random sets of patterns. It is observed that for both these sets of patterns learning in the system erases the critical state after a short initial period. Interestingly from Fig 4.6 analysing the avalanche dynamics for this case, a short duration of activity is observed, where in the system remains in the sub-critical state the system. Following this there is an almost sudden jump into a highly structured shape. Analysing the shape of this highly structured region of activity, synchronous wave like formations (bowl shaped) are observed. Looking at the mean deviation curve in Fig. 4.7 it is established that criticality is lost in the network with $\Delta \gamma$ being well above the fixed threshold of 0.005. Intuitively the wave like formations could be a result of overlaps among the patterns and rapid transition from one stored pattern to another. In order to analyse this further we use the evolution of the overlaps of memory states at different time steps (Fig 4.8).

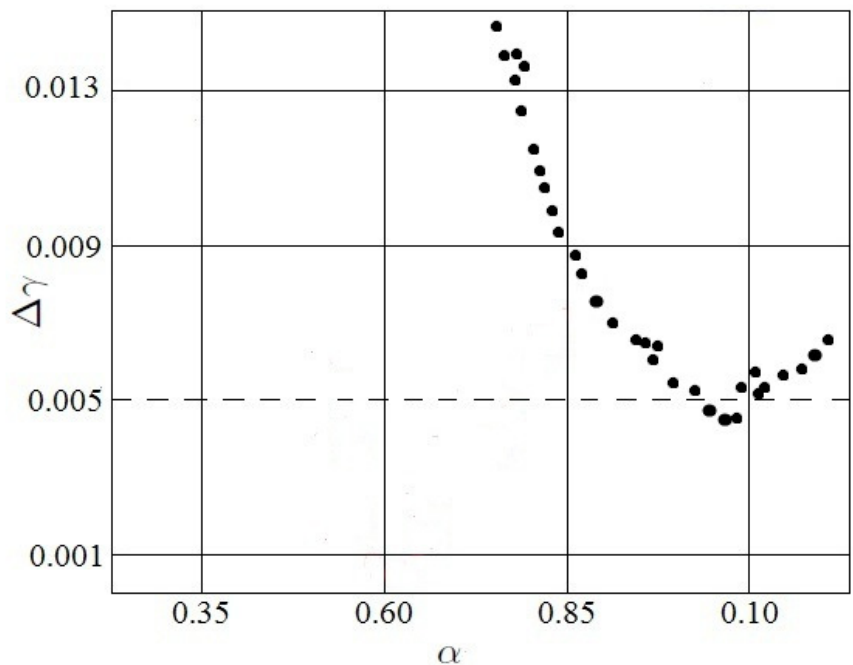

*Figure 4.7*: *The mean squared deviation from the best matching power-law is plotted in relation to the changing $\alpha$ value. The threshold describing a very good fit to the power-law distribution is fixed at 0.005. Clearly the deviation from the nearest power-law is quite high and $\Delta \gamma > 0.005$ for the observed curve establishing that the distribution is not critical.*

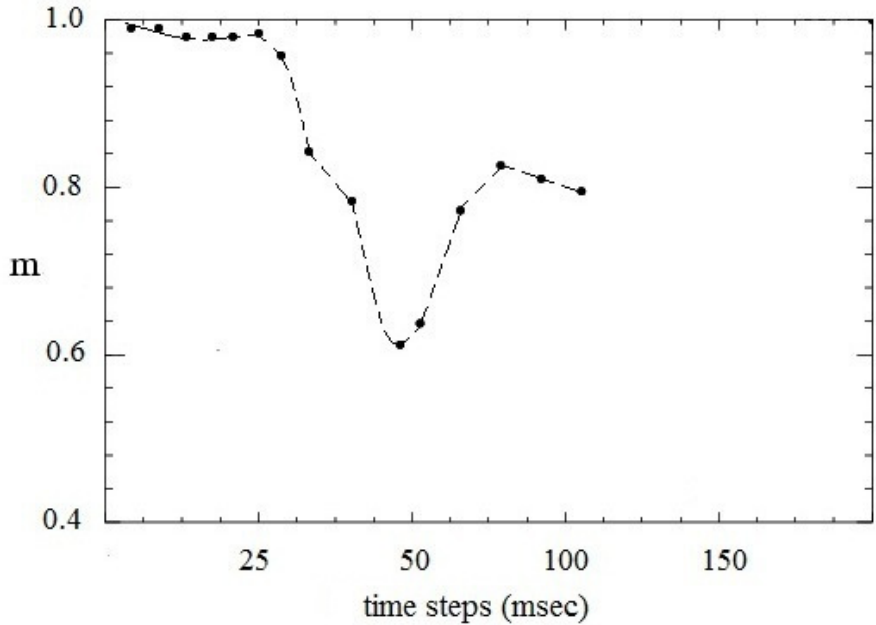

*Figure 4.8*: *Overlaps of the memory states at different time steps of the system dynamics. Simulations with N=200, P=20, 1000 trials each. The retrieval quality of the 20 random patterns stored is seen to exhibit a rather unusual pattern. Initially state coincides extremely well with one of the stored patterns. There is however a steady decline of m-value. The v-shaped curve in the figure should correspond to an attractor state denoting high degree of overlap between the stored patterns. This results in switching between patterns and the eventual rise in retrieval quality. Over time this behaviour can be believed to continue and this relating with the observed loss of criticality in the previous figure.*

The step like formation for the retrieval quality of stored patterns in Fig. 4.8 establishes that although initially there is a perfect overlap between the stored patterns and the ones retrieved (*m=1*), as the system evolves (learning process) there is an increase in the overlap between patterns. Thus we see a switching between the stored patterns which directly result in the highly structured behaviour observed during the avalanche dynamics, thus the criticality in the system is lost. E.g. for a P = 5 patterns in the network , the network first recalls pattern 5 and its anti-pattern, then switches to pattern 3 and its anti-pattern and then moves to pattern 4 and so on, frequently switching between the stored patterns.

## 4.2.3 Orthogonal patterns

However the learning of orthogonal patterns i.e. $\frac{1}{Nf(1-f)}\sum_{i=1}^{N}(\xi_i^\mu - f)(\xi_i^\nu - f) = 0$, When $\mu \neq \nu$, yields very contrary results as compared to the normal multiple pattern scenario. In this case the criticality of the system is not destroyed due to learning. The mean deviation from the power-law as a function of $\alpha$ although shows similar statistics as the single pattern case (Fig. 4.9), however it remains in the critical regime for a much shorter duration of time. This can also be observed from the plot of distribution of the avalanche sizes with varying $\alpha$ (Fig 4.10)

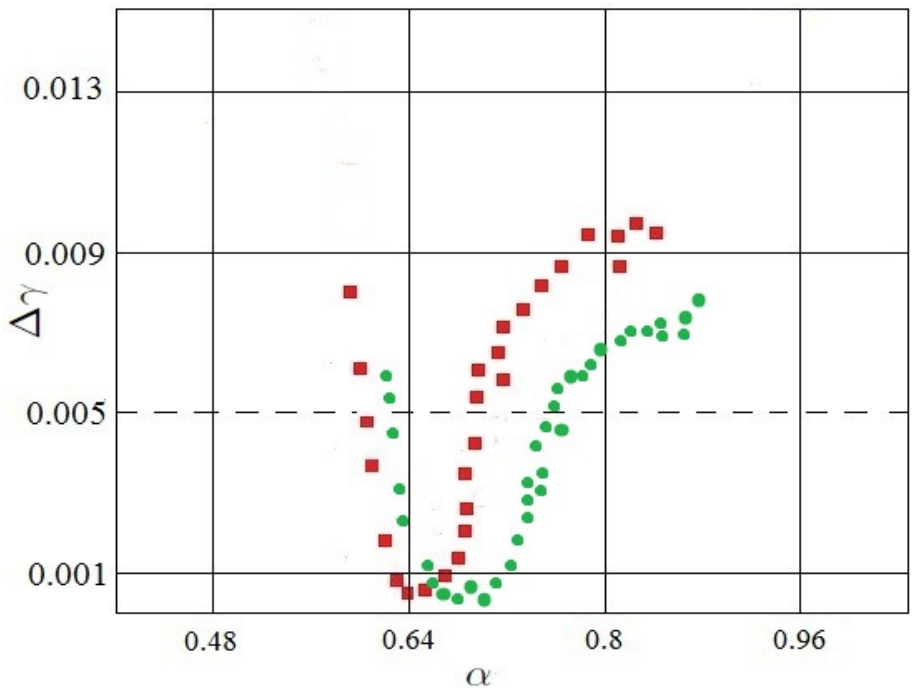

*Figure 4.9*: *The range of connectivity parameters where the critical events extend to the system size. The mean squared deviation $\Delta\gamma$ from the best matching power-law is plotted in dependence of $\alpha$. The red squares, green circles stand for networks with depressive and facilitative synapses and system sizes N=300,500, respectively. Threshold value is fixed at 0.005, same as the previous cases. Clearly there is an increase in the size of the critical region with an increase in the network size, along with a slight shift in the critical value as well. This indicates that in the thermodynamical limit the system is critical for a substantial portion.*

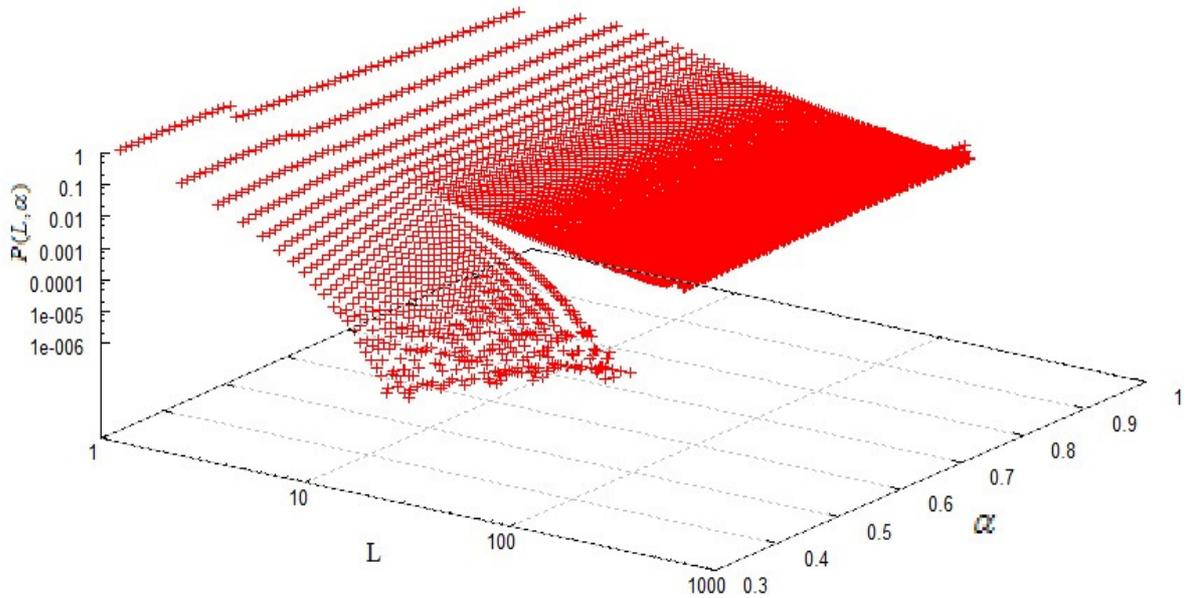

*Figure 4.10*: *The observed avalanche distribution of the orthogonal patterns case for varying $\alpha$ values. Once again a distinct change from the subcritical regime to the critical regime is visible near $\alpha = 0.64$ with a clear jump in the dynamics as a result of the facilitation in the synapses. However the critical region is visibly smaller but sharper compared to the single pattern case.*

As mentioned previously the distribution is critical if the deviation from the nearest power-law is small enough, i.e. $\Delta\gamma < 0.005$. It is observed that with the exception of the multiple pattern scenarios, all the other three special cases show that the system exhibits criticality, which is again confirmed by the plot of their corresponding avalanche distributions. However it is evident that the critical region for the orthogonal case is smaller than the critical region corresponding to the single pattern as well as the case of homeostatic regulation. Furthermore we tested the change in the shape of the mean squared deviation from bet matching power law ($\Delta\gamma$) showing that the results obtained are not restricted to any particular size of network, rather there is a growth in the size of the critical region with an increase in the number of neurons in the network, suggesting that in the thermodynamical limit (biologically realistic network sizes) the system should remain in the critical state for a substantial portion.

## 4.3 Effect of Homeostatic regulation on Criticality

In **section 3.3** we mentioned that it was possible to have synaptic adaptation outside of the Hebbian learning in the form of homeostatic plasticity that could counter balance Hebbian plasticity. Furthermore unlike the models considered until now, realistic neurons are affected by some amount of leakage that can lead to the forgetting of previously occurred synaptic input and delay a spike to occur. Hence it is essential to understand the effect of leakage on criticality. In this regard we believe that some form of homeostatic regulation may be the answer to how criticality and leakage go together in biologically realistic systems.

Here we evaluate our claim that Homeostatic regulation in the form of synaptic scaling as introduced by equation (**3.6**) ensures criticality. We use a neural memory model of N-single integrate and fire neurons, with a network size of 200 neurons that allows for associative recall, along with our introduced homeostatic regulation of the synaptic weights $W_{i,j}$. The dynamics of the system is as specified in Model introduced in section 3.2.3. For this simulation we used 10 random patterns with 50 neurons each.

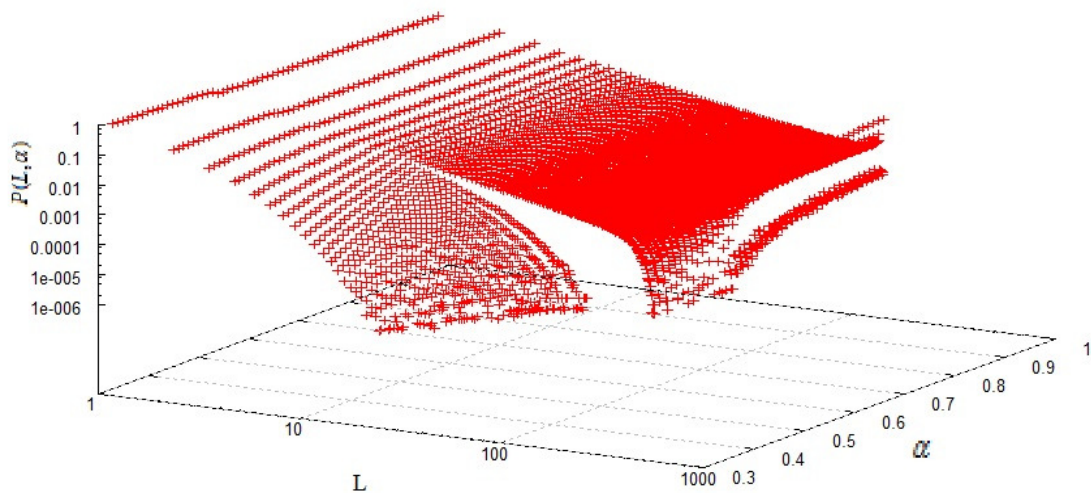

*Figure 4.12*: *Distribution of avalanche sizes for a network with synaptic adaptation beyond Hebbian learning, for different values of $\alpha$. Power-law distribution is observed with transition from sub-critical to critical occurring at the $\alpha = 0.55$. To determine the*

*critical distribution the threshold for the least squares deviation was set at 0.005. Qualitative behaviour is similar to the results of the depressive-facilitation model of [25].*

Fig 4.12 shows the avalanche duration distribution for different maximal synaptic strength values. The critical region is obtained by the condition that the mean squared deviation from the best matching power-law is not larger than 0.005. In the previous multi-pattern case although we saw that the critical region was destroyed and a more structured pattern emerges for the avalanche distribution (Fig. 4.6), in this case contrastingly the criticality is preserved. This is a direct result of the synaptic modification rule in equation (3.6) i.e.

$$\vec{W}_{i,j} = \vec{W}_{j,i} = \sum_{j=1}^{N} k_o . (\alpha - W_{i,j}) . \delta(t - t_{sp}^{j})$$

Where in, after the occurrence of a spike as the amount of neuro-transmitter approaches the maximum synaptic strength, the synaptic strength weights $W_{i,j}$ between two neurons is regulated to move towards the critical value $\alpha$. Hence there is a counter balancing of the Hebbian plasticity by this time of synaptic adaptation beyond standard Hebbian learning. The critical region is comparable to the results obtained by the original facilitation-depression model as described in Section 3.2.1 with narrow width of the $\triangle \gamma$ curve. However as seen previously in Fig. 4.9 with an increase in the network size the size of the critical region also increases, proving that the observed behaviour is consistent and not an one off recording.

## 4.4 Discussion

The results clearly point out to the fact that depending on the pattern set the synaptic adaptation is seen to cause a modulation or even a disappearance of the initial critical state. Analytically speaking and also looking at the stored pattern retrieval quality graphs, of the generic case of multiple patterns, this could be explained due to the formation of densely connected clusters within the network corresponding to the learned patterns. The destruction of criticality in the case of multiple patterns clearly helps in establishing our original hypothesis that the critical state which is the optimum for speed and comprehensive trade-off in the exploration of a large set of combinations of features, gets affected negatively as result of prolonged learning which is an integral part of the *ageing process*. It can be analysed that waves of avalanche like activity tend to reside inside the densely connected clusters, thus leading to a reduction in the exploratory effect of the network dynamics.

However the single pattern case and more interestingly the case of the orthogonal patterns demonstrate the possibility of retrieval of patterns stored in the early phase of learning. Thus highlighting the ever so important fact that the speed and extent of the loss of criticality depends on the properties of the connectivity scheme the network evolves into during the learning process. Of course one may counter question, *if there is self-organised criticality, why there are non-critical firing patterns observable?* The answer would be that this is a consequence of the structure present. There are structures like the well tuned synfire structures or chains resulting in realistic avalanche distributions in Fukai's (2007) [44] paper and perhaps the orthogonal patterns that permit criticality, but others are counteracting and could eventually be a pointer towards the wiring of real neurons. Although in general one would assume that any connectivity that induces attractors prevents criticality, but the one-pattern case as well as the orthogonal patterns (possibly also some other types of hierarchies) would contradict to this. Furthermore the use of synaptic adaptation outside of the Hebbian dynamics in the form of homeostatic regulation of the synaptic weights such that criticality is maintained could be an indication of biologically realistic mechanism in the brain to achieve optimal state of performance even in unusually complex scenarios, and preventing the development of prolonged supercritical or subcritical regimes leading to the onset of eventual epileptic seizures in the brain.

Although it is still very early to comment on the organisation of memory in the brain and the way learning effects it in biologically realistic situations, the fact that the learning of orthogonal patterns delays the loss of criticality, could open up windows for understanding of the hierarchical organisation of memory in the brain, and the effect ageing has on it leading to the cognitive deficits which are commonly attribute to old age.

One of the important reasons for orthogonality is that it enables more patterns to be stored in the network. Also, if two patterns are similar, we might be able to substitute one for the other without too much loss. Thus if we cannot store two patterns as distinct, the next best thing is to store them as one and classify them together. More generally, when there are highly correlated states, we store one prototype that could be one of the states, or even a spurious state that has maximal overlap with the correlated states. This best utilization strategy enables each subdivision to retain states that are well representative, even though not necessarily exact reproductions, of the possible states. This suggests the possibility of a hierarchical orthogonal organisation aiding in learning and stabilizing criticality such that, although prolonged learning leads to impaired performance in unusual complex situations (*fluid intelligence*), there is a possibility of making *experienced* guesses, or the commonly attributed phenomenon of "*getting wiser*" with old age can be observed due to the life-long optimization of memory patterns.

# Chapter 5

# Conclusion

In this thesis we studied the cognitive effects of the ageing brain, specifically in the context of a longer learning history using models of neural dynamics and learning, capable of associative recall based on the underlying principles of Hebbian learning and self-organised criticality. We were mainly interested in the following two ideas in the context of the ageing process:

- A possible combination of Hebbian learning scheme with the phenomenon of Self organised criticality and their eventual outcome.
- The effect of prolonged learning on the criticality of the network

We started by motivating the idea of a visible decline in the cognitive functionalities as a function of the ageing process of the brain, as observed by performances in numerous memory tasks and hence the need for sophisticated neuronal models to account for ageing in correspondence with known physiological changes in the brain. Contrary to common compartmentalised perception of the brain, we introduced the brain as a complex dynamic system and its inherent mechanisms favouring Self-organization. In chapter 2 we introduced the concepts of fluid and crystalline intelligence in order to lay the foundation for the effect of learning on memory consolidation. The standard Hopfield model along with the generic associative memory problem were introduced and discussed in length along with its drawbacks and the improvements thus considered in this project. It is believed that the critical state makes provision for optimality in the brain and in this regard we discussed the phenomenon of Self-organised criticality in detail, highlighting the analytical as well as experimental observation of critical dynamics in networks with dynamical synapses.

As a first step we implemented the neural model with dynamical synapses as originally proposed by Levia et. al [4, 25] based on the approach of Tsodyks and Makram (1996) for a reliable reproduction of synaptic responses between pyramidal neurons. A mathematical model of this network was presented along with the addition of an optimal Hebbian learning rule thus enabling us to explore the idea of associative memory with dynamical synapses using avalanche dynamics, in the light of maintaining a stable critical regime in the system.

Although the networks considered are idealised, we explored the concept of realistic leaky neurons by introducing homeostatic regulation in our model in the form of synaptic adaptation of the weights outside of Hebbian learning. Various forms of homeostatic plasticity are believed to balance Hebbian plasticity and therefore stabilize the properties of the neural circuits. It is advocated that this same homeostatic mechanisms could be responsible for ensuring criticality in the brain and thus preventing predominance of supercritical regimes as a result of the negative effect prolonged learning has on the critical state. Furthermore as a possible mechanism, adaptation of the synaptic weights $W_{i,j}$ such that the average synaptic strength remains at the critical level is introduced into the modified neural model of Hebbian learning and criticality.

Using mean squared deviation from best matching power law and the measure of the degree of overlap between memory states of stored and retrieved patterns, simulations of these neural models are carried out and evaluated. Interestingly the results obtained point towards a modulation or eventual disappearance of criticality depending on the pattern sets that are part of the synaptic adaptation. The behaviour of the random multiple patterns clearly established our primary hypothesis that, due to prolonged learning the criticality in the system is lost thus accounting for the deficit of cognitive functionalities, specifically the ability to make sense out of completely new complex situations during ageing. However it was observed that in the case of a single fixed pattern and more importantly orthogonal patterns the critical state is not lost. This could be an indication towards the evolution of the network into a connectivity scheme that maintains criticality. A considerable amount of research has been carried out by researchers in both cognitive neuroscience as well as computational neuroscience in order to explore the idea of age related cognitive effects being caused by an optimization process. We believe the results thus obtained, specifically for the orthogonal patterns case point towards such a optimization process in the brain where in the connectivity scheme evolves in a way to maintain criticality with an eventual trade off between different competing features like working memory, speed of processing, selective retention etc leading to the gradual decline in fluid intelligence and a lifelong improvement in crystalline intelligence.

All in all we believe this is another important contribution in the vast area of neuronal modelling and understanding with regards to the conjunction of the concept of self-organised criticality and the ageing process. Furthermore it is a well established fact that long term memory loss or progressive gradual deterioration is a characteristic phenomenon of age related disorders like Alzheimer's.

Although at a very nascent stage and still far from biological implications, we believe as a future direction our results could be used to understand the deterioration of memory retrieval in a greater light, opening up possibilities to model and explore the organization and evolution of memory in the brain and thus helping researchers in understanding the complexities of these age related disorders.

# Appendix A

## A.1 Power-law distributions

A power-law is a special kind of mathematical relationship between two quantities such that when the frequency of an object or event varies as a power of some attribute of that object or event, the frequency is said to follow a power-law.

In general a power-law distribution is any that has the following form:

$$p(x) \propto g(x) x^{-\alpha} \qquad (A.1)$$

Where, $\alpha > 1$ and $g(x)$ is any function that satisfies $\lim_{x \to \infty} \frac{g(tx)}{g(x)} = 1$ with $t$ a constant.

Furthermore since or a power-law it is required that $p(x)$ be asymptotically scale invariant the form of $g(x)$ only controls the shape and finite extent of the lower tail. For instance, if $g(x)$ is a constant function, then we have a power-law that holds for all values of $x$. In many cases, it is convenient to assume a lower bound $x_{min}$ from which the law holds. Combining these two cases, and where $x$ is a continuous variable, the power law has the form $p(x) = \frac{\alpha - 1}{x_{min}} (\frac{x}{x_{min}})^{-\alpha}$, where the pre-factor to $x^{-\alpha}$ is the normalizing constant.

In general, power-law distributions are plotted on doubly logarithmic axes, which emphasize the upper tail region. The most convenient way to do this is via the cumulative distribution $p(x) = \Pr(X > x)$, Where

$$\Pr(X > x) = C \int_x^\infty p(X) dX = \frac{\alpha - 1}{x_{min}^{-\alpha+1}} \int_x^\infty X^{-\alpha} dX = \left(\frac{x}{x_{min}}\right)^{-\alpha+1} \qquad (A.2)$$

Power-law distributions have been commonly observed in a variety of studies from the populations in cities all over the world, word frequencies in literature, the strength of earthquakes, the wealth of individuals, forest fires, distinction of biological species, to even the distribution of web site pages. However although power-laws seem to be an inherent part of nature, their dynamical origin is still not very clear.

## A.2 Maximum Likelihood estimation of Power-law

In real world scenarios more than often power-laws are present as in equation A.1, however with an exponential cut-off being present, given by.

$$p(x) = C_{\gamma,\beta} x^{-\gamma} e^{-\beta x} \qquad (A.3)$$

Where, $C_{\gamma,\beta} = 1/\int_0^\infty x^{-\gamma} e^{-\beta x} dx$ is the normalization constant.

It is in these cases that linear regression provides the best estimate based on a small number of observables. Since in the case of neuronal avalanches a sharp cut-off is observed, linear regression estimation technique is used to determine the power-law parameters. Furthermore, since the parameter estimated by linear regression are the same as the parameters obtained by Maximum likelihood estimation (MLE) we discuss the later here in detail, with the primary goal being the estimation of the power-law exponent $\gamma$.

Experimental results can assumed to be statistically independent and therefore we can represent the data $x_1.........x_n$ in $n$ measurements by,

$$L(\gamma) = \prod_{i=1}^{n} P(x_i \mid \gamma) \qquad (A.4)$$

Where, $L$ is the likelihood function.

Taking logs on both sides in-order to avoid dealing with extreme values, we get the following logarithmic likelihood.

$$\ln L(\gamma) = \prod_{i=1}^{n} \ln P(x_i \mid \gamma) \qquad (A.5)$$

We introduce $\hat{\gamma}$ the value that maximizes $L$, known as the maximum likelihood estimate for the real $\gamma$. We obtain $\hat{\gamma}$ by solving,

$$\frac{\partial (\ln L(\gamma))}{\partial \gamma} = 0 \qquad (A.6)$$

Now, for a continuous power-law distribution the density is given by,

$$p(x) = (\gamma-1)x^{\gamma} \qquad (A.7)$$

Therefore the logarithmic likelihood function takes the following form,

$$\ln L(\gamma) = n\ln(\gamma-1) - \gamma\sum_{i=1}^{n}\ln x_i \qquad (A.8)$$

Hence solving A.6 we get the following estimate of the power-law exponent,

$$\hat{\gamma} = 1 + n\left(\sum_{i=1}^{n}\ln x_i\right)^{-1} \qquad (A.9)$$

For the discrete power-law distribution normalization constant is given by Riemann zeta function

$$C_\gamma = \frac{1}{\zeta(\gamma)} = \frac{1}{\sum_{k=1}^{\infty}k^{-\gamma}} \qquad (A.10)$$

Therefore analogous to equations **(A.7),(A.8),(A.9)** one can find the estimate for the exponent as solution of the equation,

$$-\sum_{i=1}^{n}\ln x_i = \eta\frac{\zeta'(\gamma)}{\zeta(\gamma)} \qquad (A.11)$$

However it is possible to find solution to equation **(A.11)** only numerically. Furthermore since the avalanche size distribution is discrete we either use this estimation or explicitly look for the maximization of the logarithmic likelihood function.

# Bibliography

[1]     P. Bak, C. Tang and K. Wiesenfield, Self-Organized criticality: An Explanation of the 1/f noise. *Phys. Rev. Lett*. 59, 381-384, (1987)

[2]     J. Beggs and D. Plenz, Neuronal avalanches in neocortical circuits. *J. Neurosci*. 23, 11167-11177, (2003)

[3]     D. Chialvo, Are our senses critical?, *Nature Physics* 2, 301-302, (2006)

[4]     C.W Eurich, M.Herrmann and U.Ernst, Finite-size effects of avalanche dynamics. *Phy. Rev*. E 66, 066137-1-15, (2002)

[5]     A. Levina, M.Herrmann and T. Geisel, Dynamical Synapses causing Self-Organised criticality in neural networks, *Nature Physics*, 3, 857-860, (2007)

[6]     B. Rypma, M.D'Esposito, Isolating the neural mechanisms of age-related changes in Human working memory. *Nature Neuroscience*, Vol. 3, No.5, 509-515. (2000)

[7]     T.A. Salthouse, The Processing-speed theory of adult age differences in cognition. *Phychol Rev*. 103(3), 403-428. (1996)

[8]     V.F. Shefer, Absolute number of neurons and thickness of the cerebral cortex during ageing, secile and vascular dementia and Pick's and Alzheimer's disease. *Neuroscience and Behavioural physiology*. Vol. 6, No. 4, 319-324 (1973)

[9]     Horn, JL and Cattell, RB. Age differences in fluid and crystallized intelligence. *Acta Psychologica*. 107, (1967)

[10]    Braver, T.S and Barch, D.M. A theory of cognitive control, ageing cognition, and neuromodulation. *Neuroscience and Biobehavioural Reviews* 809-817. (2002)

[11]    Brown, J.W. , Reynolds, J.R and Braver, T.S. A computational model of fractionated conflict-control mechanisms in fast-switching *cognitive psychology* 37-85. (2007)

[12]    Li, S.C and von Oertzen, T. And Lindenberger, U. A neurocomputational model of stochastic resonance and ageing. *Neuro-computing* 1553-1560, (2006)

[13]    Nadel, L., Moscovich, M. Memory consolidation, retrograde amnesia and the hippocampal complex. *Curr Opin. Neurobiol*. 7:217227, (1997)

[14]    Hopfield, J.J., Neural networks and physical systems with emergent collective computational abilities. *Proc. of National Academy of Sciences*, Vol. 79, No. 8, 2554-2558, (1982)

[15]    Chialvo, D.R., Critical Brain networks. *Physica A* 340(4):756, (2004) *Contribution of the Neils Bohr summer institute on complexity and criticality* (2003); *in Per Bak memorial issue of Physica A.*

[16]    V. Frette, K. Christensen, A.M. Malthe-Sorenssen, J. Feder, T. Jossang, and P.Meakin. Avalanche dynamics in a pile of rice. *Nature* 397, 49-52, (1996).

[17]  B. Gutenberg and C.F. Richter. Magnitude and energy of earthquakes. *Annali di Geofisica* 9, 1-15, (1956).

[18]  Herz, Andreas V.M., and Hopfield, John J., Earthquake cycles and Neural Reveberations: Collective Oscillations in Systems with Pulse-Coupled Threshold Elements. *Phy Rev. Lett,* Vol. 75, No. 6, 1222-1225, (1995).

[19]  H.J.S. Feder and J.Feder. Self-Organized criticality in a stick-slip process. *Phy. Rev Lett.* 66, 2669-2672, (1991).

[20]  R. A. Legenstein, W. Mass. Edge of chaos and prediction of computational performance for neural microcircuit models. *Neural Networks*, 323-333 (2007).

[21]  C. Haldeman and J. Beggs. Critical branching captures activity in living neural networks and maximizes the number of metastable states. *Phys. Rev Lett.* 94, 058101, (2005).

[22]  J. Begs and D.Plenz. Neuronal avalanches are diverse and precise activity patterns that are stable for many hours in cortical slice cultures. *J. Neuroscience*, 24:22, 5216 – 5229, (2004).

[23]  D. Plenz, T.C. Thiagarajan. The organizing principles of neuronal avalanches : Cell assemblies in the cortex ? *Trends in Neurociences,* 30, 3 ,101-110, (2007).

[24]  D. Horn, E. Ruppin, M. Usher, M. Herrmann, Neural network modeling of Memory Deterioration in Alzheimer's disease. *Neural Computaion*, Vol. 5, No. 5, 736-749, (2008).

[25]  A. Levina, M.Herrmann and T. Geisel, Phase transitions towards Criticality in a Neural System with adaptive interactions, *Physics Review letters*, 102, 118110, (2009)

[26]  M. Tsodyks, K. Pawelzik and H. Markram, Neural networks with Dynamic Synapses,*Neural Computation*, Vol. 10, No. 4, 821-835, (1998)

[27]  L. Pantic, J. Torres, H. Kappen, S. Gielen, Associative memory with dynamic synapses, *Neural Computation,* Vol. 14, No. 12, 2903-2923, (2002)

[28]  G. Turrigiano, Homeostatic plasticity in neuronal networks: the more things change, the more they stay the same, *Trends Neurosci.,*22, 221-227, (1999)

[29]  G. Turrigiano, S. Nelson, Hebb and homeostasis in neuronal plasticity, *Current opinion in Neurobiology*, 10, 358-364, (2000)

[30]  G. Turrigiano, L. Abbott and E. Marder, Activity changes the intrinsic properties of cultured neurons, *Science*, 264, 974-976, (1994)

[31]  D. Bibitchkov, J. M. Herrmann and T. Geisel, Pattern storage and processing in attractor networks with short-time synaptic dynamics, *Network: Computation in neural systems*, Vol. 13, No. 1, 115-129, (2002)

[32]  D.J. Amit, H. Gutfrend and H. Sompolinsky, Information storage in neural networks with low levels of activity, *Physical review*, Vol. 35, No. 5, 2293-2303, (1987)

[33]  N. Peng, N.K. Gupta and A.F. Armitage, An investigation into the improvement of local minima of the Hopfield network, *Neural networks*, (1996)

[34] D. Wilshaw and P. Dayan, Optimal plasticity from matrix memories: what goes up must come down, *Neural Computation* , (1990)

[35] G. Checkik, I. Meilijsm and E. Ruppin, Effective neuronal learning with ineffective Hebbian learning rules, *Neural Computation*, (2001)

[36] P. Bak. How nature worksL the Science of Self-organised criticality. *Springer Verlag.* (1999)

[37] G. Buzsaki, Rythms of the brain. *Oxford University press*. (2006)

[38] M.E.J. Newman, Evidence for Self-organised criticality in evolution, *Physica D*, 107:293296 (1997)

[39] B. A. Pearlmutter, C. J. Houghton and S. Tilbuiy , A new hypothesis for sleep: Tuning for criticality, *Neural Computation*, (2009)

[40] W. Senn, I. Seger, and M. Tsodyks, Reading neuronal synchrony with depressing synapses, *Neural Computation*, 10, 815-819, (1998)

[41] A. Levina, A mathematical approach to Self-Organised Criticality in Neural networks, *PhD dissertation, Gottingen,* (2008)

[42] L. de. Arcangelis, C. Perrone-Capano and H. J. Herrmann, Self-organised criticality model for brain plasticity, *Phys Review Lett.*, 96: 028107(4)

[43] G.L. Pellignini, L. de. Arcangelis, H. J. Herrmann and C. Perrone-Capano, Activity-dependent neural network model on scale free networks, *Physical Review,* 76(1), (2007)

[44] J. Teramae and T. Fukai, Local cortical circuit model inferred from power-law distributed neuronal avalanches, *Journal of computational neuroscience,* Vol. 22, No. 3, 301-312, (2007)

[45] M. Daneman and P. A. Carpenter, Individual differences in working memory and reading, *Journal of verbal learning and verbal behaviour,* Vol. 19, No. 4, 450-466, (1980)

[46] J. Hertz, K. Anders and R.G. Palmer, Introduction to the theory of neural computation, *A Lecture notes volume in the Santa Fe Institute studies in the sciences of complexity.* (1991)

[47] G. Turrigiano, K.R. Leslie, N.S. Desai, L.C. Rutherford and S.B.Nelson, Activity dependent scaling of quantal amplitude in neocortical neurons, *Nature*, (1998)

[48] L.E Dobrunz and C.F. Stevens, Heterogeneity of release probability, facilitation and depletion at central synapses, *Neuron*, 995-1008

[49] M. V, Tsodyks and H. Markram, The neural code between neocortical pyramidal neurons depends on neurotransmitter release probabability. *PNAS*, 94, 719-723 (1997)

[50] A. M. Thomson and J. Deuchar, Temporal and spatial properties of local circuits in neo-cortex. *Trends Neurosci.*, 17:119-136 (1994)